\documentclass[final,3p,times]{elsarticle}
\usepackage{amssymb}
\usepackage[cmex10]{amsmath}

\usepackage[section]{placeins} % 

\journal{Journal of Sound and Vibration}

\usepackage{color}
\newcommand{\revtext}[1]{{{\color{black}{#1}}}}
\newcommand{\revtextt}[1]{{{\color{black}{#1}}}}

\newcommand{\vddtwo}[2]{	
	\left\{ \!\!
	\begin{array}{cc}
	#1 \\ 
	#2 
	\end{array}
	\!\! \right\}
	}

\renewcommand{\imath}{i} 
\newcommand{\Kappa}{k_c} 
\newcommand{\qwave}{\kappa}

\begin{document}

\begin{frontmatter}



\title{Supratransmission in a Disordered Nonlinear Periodic Structure}


\author[dal]{B.~Yousefzadeh\corref{cor1}}
\ead{behroozy@alumni.ubc.ca}

\author[dal]{A.~Srikantha~Phani}
\ead{srikanth@mech.ubc.ca}

\cortext[cor1]{Corresponding author}

\address[dal]{Department of Mechanical Engineering, University of British Columbia, \\ 2054 - 6250 Applied Science Lane, Vancouver, BC, Canada, V6T 1Z4}

\begin{abstract}
We study the interaction \revtext{among} dispersion, nonlinearity, and disorder effects in the context of wave transmission through a discrete periodic structure, subjected to continuous harmonic excitation in its stop band. 
We consider a damped nonlinear periodic structure of finite length  with disorder. Disorder is introduced throughout the structure by small  changes in the stiffness parameters drawn from a uniform statistical distribution.  
Dispersion effects  forbid wave transmission within the stop band of the linear periodic structure.  However, nonlinearity leads to  supratransmission  phenomenon, by which enhanced wave transmission occurs within the stop band of the periodic structure when forced at an amplitude exceeding a certain threshold. 
The frequency components of the transmitted waves lie within the pass band of the linear structure, where  disorder is known to cause Anderson localization. There is therefore a competition between dispersion, nonlinearity, and disorder in the context of supratransmission.  
We show that supratransmission persists in the presence of disorder. The influence of disorder decreases in general as the forcing frequency moves away from the pass band edge, reminiscent of dispersion effects subsuming disorder effects in linear periodic structures. We compute the dependence of the supratransmission force threshold on nonlinearity and strength of coupling between units. 
We observe that nonlinear forces are confined to the driven unit for weakly coupled systems. This observation, together with the truncation of higher-order nonlinear terms, permits us to develop closed-form expressions for the supratransmission force threshold. In sum, in the frequency range studied here, disorder does not influence the supratransmission force threshold in the ensemble-average sense, but it does reduce the average transmitted wave energy.

\end{abstract}


\end{frontmatter}

%\linenumbers

% \tableofcnotents
%\newpage

%%%%%%%%%%%%%%%%%%%%%%%%%%%%%%%%%%%%%%%%%%%%%%%%%%%%%%%%%%%%%%%%%%%%%%%%%%%%%%%%%%%%
%%%%%%%%%%%%%%%%%%%%%%%%%%%%%%%%%%%%%%%%%%%%%%%%%%%%%%%%%%%%%%%%%%%%%%%%%%%%%%%%%%%%
%%%%%%%%%%%%%%%%%%%%%%%%%%%%%%%%%%%%%%%%%%%%%%%%%%%%%%%%%%%%%%%%%%%%%%%%%%%%%%%%%%%%
%%%%%%%%%%%%%%%%%%%%%%%%%%%%%%%%%%%%%%%%%%%%%%%%%%%%%%%%%%%%%%%%%%%%%%%%%%%%%%%%%%%%
%%%%%%%%%%%%%%%%%%%%%%%%%%%%%%%%%%%%%%%%%%%%%%%%%%%%%%%%%%%%%%%%%%%%%%%%%%%%%%%%%%%%
%%%%%%%%%%%%%%%%%%%%%%%%%%%%%%%%%%%%%%%%%%%%%%%%%%%%%%%%%%%%%%%%%%%%%%%%%%%%%%%%%%%%
\section{Introduction}
\label{sec:intro}

Spatially periodic structures exhibit intriguing dynamic characteristics, contributing to their diverse (and fast-growing) applications as phononic crystals and acoustic metamaterials~\cite{husseinAMR}, as well as light weight lattice materials~\cite{SPjasa}. 
Within the linear operating range (i.e. for motions with small amplitudes), the dynamic response of periodic structures is very well understood \cite{brillouin}. 
Of particular interest in engineering applications is the wave-filtering characteristics of one-dimensional periodic structures: for an exactly-periodic structure with no damping, the structure acts as a band-pass filter. Waves with frequency components within particular intervals (known as pass bands) travel through the structure unattenuated, whereas all other frequency components are spatially attenuated as they propagate through the structure.

For a linear, infinitely-long, exactly-periodic structure with no damping, the attenuation caused by periodicity occurs because of the destructive interference of waves that are scattered due to periodic changes in the impedance of the medium -- this phenomenon is known as \emph{Bragg scattering}. The corresponding decay is exponential and uniform at each unit. \revtext{Furthermore, the decay rate increases as the forcing frequency moves farther into the stop band.}
Periodic structures that are relevant in engineering applications have common features that further influence their dynamic response. These features can be divided into four main categories: (i) energy dissipation, (ii) deviations from exact periodicity, (iii) nonlinearity, and (iv) finite length of the structure. 
\begin{enumerate}[(i)]
	\item Addition of dissipative forces results in a new attenuation mechanism for the traveling waves. For a linear viscous dissipative force, the corresponding decay rate is exponential and uniform, and could influence all frequencies. 
	In addition, dissipative forces (if they are strong enough) can result in appearance of \emph{spatial} stop bands, corresponding to prohibited wavenumbers \cite{husseinDamping,PhaniHussein}.
	\item Small deviations from exact periodicity may lead to qualitatively different dynamic response under certain conditions \cite{defect,Hodges82}. These deviations could be a result of either \emph{defect} or \emph{disorder}. 
	A defect is normally a large deviation from periodicity that is concentrated at a certain location (or locations) within the structure, such as a crack. Disorder usually refers to small deviations from exact periodicity spread \emph{throughout} the structure, such as those due to unavoidable manufacturing tolerances. 
	\item As the amplitudes of motion increase, nonlinear forces in the structure may become significant. Nonlinearity can result in changes in location and/or width of the pass band \cite{vakakisJASA,chakraborty2001}, \revtextt{offering a mechanism to tune} the filtering characteristics. 
	Moreover, nonlinearity can lead to a new energy transmission mechanism within the stop band of a periodic structure, which \revtextt{is not possible} in the linear regime~\cite{supra_00,supra_00_exp}. 
	\item Finite length of a structure results in reflection of waves from its boundaries. This results in formation of standing waves with restricted wavenumbers~\cite{sengupta}. 
\end{enumerate}

{Influence} of some of the above four factors on the dynamic response of periodic structures have been studied in the literature. 
The presence of defects leads to localization of response near the defect \cite{defect}. This feature has been utilized in nonlinear systems to develop acoustic switches \cite{daraioNature}.  
In a disordered periodic structure, the scattering of waves at each unit is different from that of the other units due to the random nature of disorder, and the mode shapes of a typical disordered structure become spatially localized. This may result in an overall exponential decay of the response amplitude away from the source of excitation, provided that adjacent units are weakly coupled together \cite{Hodges82,HodgesWoodASA}. 
This phenomenon is known as \emph{Anderson localization}, and is explained in more detail in Section \ref{sec:disordered}. 
An overview of simultaneous effects of damping and disorder in linear periodic structures can be found in \cite{CastanierIndividual}. See \cite{chaAIAA} for an extension of this work to structures with finite length. 

Nonlinear forces make it possible for a discrete periodic structure to sustain time-harmonic solutions that have a spatially localized profile. These solutions are known as \emph{discrete breathers} or \emph{intrinsic localized modes} \cite{DBgeneral}, and their study in the engineering literature was pioneered in microelectromechanical systems \cite{DBcolloq,DBmote}. Apart from this, nonlinearity offers a route to achieve enhanced transmission within the stop band of a periodic structure\revtextt{, as mentioned earlier}. 
This may happen due to instability of periodic solutions through either saddle-node bifurcation~\cite{supra_aubry,supra_susanto} or nonlinear resonances~\cite{daraioCombinationRes}. 
The former case, called \emph{supratransmission}, occurs when the periodic structure is forced at one end with a frequency within its linear stop band. For small driving force, the amplitudes of motion are small and decay exponentially away from the driven end of the structure; i.e. energy transmission is not possible. 
If the driving force is larger than a threshold, energy transmission becomes possible even though the forcing frequency remains within the stop band of the structure.

In supratransmission, the frequency components of the nonlinearly transmitted waves lie within the linear pass band of the structure. This is true for either infinite Hamiltonian systems \cite{supra_aubry} or damped systems with a small number of units \cite{JSV1}, and can have important consequences for the transmitted energies if the periodic structure is disordered. 
The reason is that disorder has its most significant influence on energy transmission for frequencies within the pass band. It is therefore natural to expect a competition between nonlinearity and disorder with regards to transmitted energies above the supratransmission threshold.
Our main goal is to study this interaction.

There is a myriad of papers on the interplay between nonlinearity and disorder in periodic structures. The majority of this literature, originating within the physics community, focuses on conservative and infinitely-long systems. In that context, the main goal is to understand the spreading process of initially-localized wave packets (an initial value problem); see~ \cite{nonlinear_disorder_2013,nonlinear_disorder_2014} for recent reviews of this topic. 
Hereafter, we only discuss relevant studies on disordered structures within the nonlinear mechanics literature; refer to Section 4.2 in \cite{husseinAMR} for a review of the corresponding literature on ordered structures. 

Some of the earlier works on nonlinear disordered structures include~\cite{Sayar,Emaci,KingLayne}. These studies deal with the free response of conservative systems; thus, their results are not applicable to the case of supratransmission. 
Many of the remaining studies are concerned with the evolution of pulses as they propagate through nonlinear disordered structures~\cite{maynard_pulse,boechlerDisorder,masonDisorder,theocharisDisorder}, particularly for granular chains~\cite{boechlerDisorder,masonDisorder,theocharisDisorder}. 
The specific decay characteristics of pulses (exponential or algebraic) is found to depend on the relative strengths of nonlinearity and disorder, as well as the specific form of nonlinearity. Unfortunately, these results are not directly applicable to the case of continuous wave excitation in supratransmission because superposition does not generally hold in nonlinear systems. 

Wave propagation due to continuous harmonic excitation has been studied in strings loaded with masses, both experimentally and numerically. 
It was reported in the experimental work \cite{Maynard_string,Maynard_all} that increasing the driving amplitude leads to a decrease in the transmitted energy within the pass band. 
The reported results, though, are obtained for only one realization of disorder. Besides, the response regime is confined to driving amplitudes below the onset of anharmonic motion, which would be below the onset of supratransmission. 
In the numerical work~\cite{richouxNonlinearAnderson}, transmitted energies were found to generally increase with the amplitude of incident waves. Although anharmonic regimes are observed at high intensities, the corresponding increases in transmitted energies are surprisingly small in these regions (especially for an undamped system). This is uncharacteristic of supratransmission.

We are aware of two main studies that directly address the phenomenon of enhanced energy transmission in nonlinear disordered structures due to continuous harmonic excitation~\cite{supra_disorder,supra_disorder_pikovsky}. 
These studies report the existence of transmission thresholds for undamped structures of long~\cite{supra_disorder} and (relatively) short~\cite{supra_disorder_pikovsky} lengths. In both studies, the excitation frequencies are limited to the pass bands of the corresponding ordered structures (this is the frequency range in which disorder prohibits energy transmission from a linear perspective). Also, energy dissipation occurs at the boundaries of the periodic structures, but there is no internal energy loss within the periodic structures themselves. 
In~\cite{supra_disorder_pikovsky}, the statistical influence of disorder on the average response was studied for different structure lengths, and it was shown that the average threshold decreases with the number of units as a power law. 
In the limit of an infinitely long structure, however, the average transmission threshold is expected to remain finite~\cite{supra_disorder}.

In contrast to the above studies~\cite{supra_disorder,supra_disorder_pikovsky}, we focus on forcing frequencies within the linear stop band of the structure; we are specifically dealing with supratransmission. In addition, our periodic structure is damped (internal energy loss) and has finite length -- these two conditions are necessary to make the results directly applicable to engineering structures. Under these conditions, our goal is to better understand the influence of disorder on supratransmission thresholds and, in particular, on the corresponding transmitted energies.
The effects of damping, type of nonlinearity and strength of coupling on supratransmission have been previously studied for an exactly-periodic structure with finite length~\cite{JSV1}. 
The present study is a continuation of~\cite{JSV1}, investigating the statistical influence of linear disorder on (i) the transmission thresholds away from the linear pass band, (ii) transmitted energies above the transmission threshold, (iii) the spectrum of the nonlinearly transmitted waves. {We also highlight the important role of damping at frequencies near a pass band, whereby the onset of supratransmission would no longer occur at the first saddle-node bifurcation point. Accordingly, we restrict our attention to frequencies far from the pass band.}

We show that supratransmission persists in this setting in the presence of disorder. More importantly, we find that, \emph{in an ensemble-average sense}, the onset of transmission is the same in ordered and disordered structures and the transmitted spectrum lies within the pass band of the underlying linear disordered structure. These results are obtained for different strengths of disorder taken from a uniform distribution.
We also provide analytical estimates for the onset of supratransmission in weakly-coupled structures; this is the coupling regime in which disorder effects are most relevant \revtextt{to} engineering applications.

Throughout the work, we distinguish between periodic structures with exact periodicity and those with disorder by referring to them as ordered and disordered structures, respectively. 
We start in Section~\ref{sec:linearperiodic} with a review of the known results for linear periodic structures. 
Nonlinear energy transmission in ordered structures is then briefly reviewed in Section~\ref{sec:nonlinearPeriodic} for excitations both within a stop band (supratransmission) and within a pass band. 
We discuss the statistical effects of linear disorder on supratransmission in Section \ref{sec:supraDisordered}. 
In Section \ref{sec:prediction}, we present an analytical formula for predicting the onset of transmission in ordered and disordered systems, and investigate its range of validity. 
This concludes the paper in Section \ref{sec:conclude} with a summary of the main findings.

%%%%%%%%%%%%%%%%%%%%%%%%%%%%%%%%%%%%%%%%%%%%%%%%%%%%%%%%%%%%%%%%%%%%%%%%%%%%%%%%%%%%
%%%%%%%%%%%%%%%%%%%%%%%%%%%%%%%%%%%%%%%%%%%%%%%%%%%%%%%%%%%%%%%%%%%%%%%%%%%%%%%%%%%%
%%%%%%%%%%%%%%%%%%%%%%%%%%%%%%%%%%%%%%%%%%%%%%%%%%%%%%%%%%%%%%%%%%%%%%%%%%%%%%%%%%%%
%%%%%%%%%%%%%%%%%%%%%%%%%%%%%%%%%%%%%%%%%%%%%%%%%%%%%%%%%%%%%%%%%%%%%%%%%%%%%%%%%%%%
%%%%%%%%%%%%%%%%%%%%%%%%%%%%%%%%%%%%%%%%%%%%%%%%%%%%%%%%%%%%%%%%%%%%%%%%%%%%%%%%%%%%
%%%%%%%%%%%%%%%%%%%%%%%%%%%%%%%%%%%%%%%%%%%%%%%%%%%%%%%%%%%%%%%%%%%%%%%%%%%%%%%%%%%%
\section{Energy Transmission in Linear Periodic Structures}
\label{sec:linearperiodic}

%%%%%%%%%%%%%%%%%%%%%%%%%%%%%%%%%%%%%%%%%%%%%%%%%%%%%%%%%%%%%%%%%%%%%%%%%%%%%%%%%%%%
%%%%%%%%%%%%%%%%%%%%%%%%%%%%%%%%%%%%%%%%%%%%%%%%%%%%%%%%%%%%%%%%%%%%%%%%%%%%%%%%%%%%
\subsection{Proposed Periodic Structure with Tunable Nonlinearity}
%\hspace{15mm}
\label{sec:repetition}

\begin{figure}[htb]
	\centering
	\includegraphics[width=0.9\linewidth]{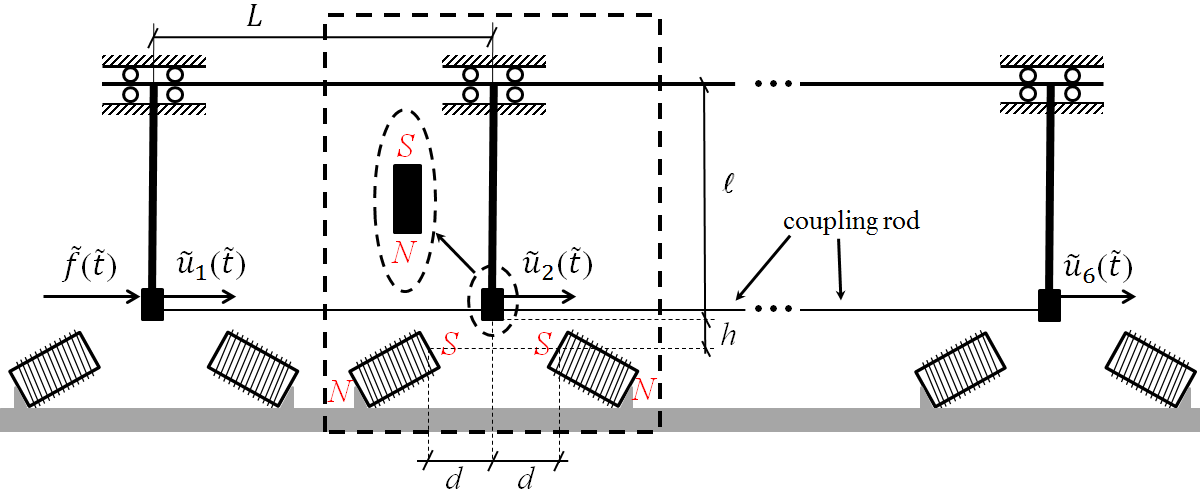}
	\caption{
	The schematic of the periodic structure made of $N$ unit cells. The repeating unit is indicated by the dashed box. The external harmonic force, $\tilde{f}(\tilde{t}\,)$, is applied to the first unit only. Two electromagnets, operated by direct currents, are fixed to the ground under the beam in each unit. The currents are chosen such that the electromagnets have the same polarity facing the beam. {We fix $h=d/10$ in this work.}
	}
	\label{fig_chain}
\end{figure}

Figure \ref{fig_chain} shows the proposed periodic structure consisting of $N$ repeating units. Each unit is made of a thin, suspended cantilever beam of length $\ell$ with a tip mass (a permanent magnet). Two electromagnets (hatched rectangles in Figure~\ref{fig_chain}) are fixed to the ground at a vertical distance $h$ below the tip mass, and interact with the permanent magnet (black rectangles). The electromagnets are symmetrically placed from the beam axis at a horizontal distance $d$. This symmetric arrangement ensures that the vertical position of the cantilever is the equilibrium configuration. 
Direct current (DC) is passed through each electromagnet such that they have the same polarity facing the beam. The magnetic forces between the permanent magnet at the tip and the two electromagnets provide tunable nonlinear restoring force for each beam.  
The first beam is excited with a harmonic force $\widetilde{f}(\tilde{t}\,)=\widetilde{F} \cos(\omega\!_f \tilde{t} \,)$, {where $\widetilde{F}$ is the magnitude of the applied force, $\omega\!_f$ is the driving frequency and $\tilde{t}$ is time.}
A coupling rod of length $L$ couples the displacements of adjacent beam. The spacing between adjacent beams ($L$) is large compared to the separation between the magnets ($2d$) in order to avoid magnetic interference effects.

%%%%%%%%%%%%%%%%%%%%%%%%%%%%%%%%%%%%%%%%%%%%%%%%%%%%%%%%%%%%%%%%%%%%%%%%%%%%%%%%%%%%
\subsection{Governing Equations for the Periodic System}
\label{sec:EOM}

We focus on energy transmission in frequency ranges within the first pass band, where all cantilever beams vibrate in their fundamental mode with different phases. Thus, the governing system of partial differential equations reduces to a set of coupled ordinary differential equations. 
In this case, the equations of motion for free vibrations of an isolated unit can be written in terms of the displacements of the tip of the beam, $\widetilde{u}_n\,(\tilde{t}\,)$, as 
\begin{equation}
	\label{EOMunit}
	\ddot{\widetilde{u}}_n + 2\widetilde{\zeta} \dot{\widetilde{u}}_n + \omega_1^2 \, \widetilde{u}_n + \widetilde{F}_{M,n} = 0
\end{equation}
where overdot denotes time derivative, $\widetilde{\zeta}$ is the coefficient of viscous damping, $\omega_1$ is the first natural frequency of the beam when the electromagnets are removed and $\widetilde{F}_M$ is the horizontal component of the magnetic force acting on the tip of the beam resulting from the interaction of the permanent magnet and two electromagnets.
If the displacements of the tip of the beam are small compared to its length, equal currents pass through the two electromagnets within the unit cell, and the magnets are modeled as magnetic poles, then the magnetic force $\widetilde{F}_{M,n}$ can be written as follows:
\begin{equation}
  \label{FMunit}
      \widetilde{F}_{M,n} =   \widetilde \mu_n 
      \frac{(d+\widetilde{u}_n)} {\left( (d+\widetilde{u}_n)^2 + h^2 \right)^{3/2}} 
             -\widetilde \mu_n 
      \frac{(d-\widetilde{u}_n)} {\left( (d-\widetilde{u}_n)^2  +h^2 \right)^{3/2}}
\end{equation}
The constants $\widetilde \mu_n$ depend on the strengths of the magnetic poles and contain the non-geometric dependencies of the magnetic force.
The magnetic force can be tuned, thus providing control over the strength of nonlinearity, as well as its type (softening or hardening).  

To normalize the governing equations, we use $\omega_1$ for time and $d$ for displacements.  Considering linear coupling between the units due to the coupling rods, accounting for the external force at the first unit, and dividing all terms by $\omega_1^2d$, we arrive at the non-dimensional form of the governing equations in \eqref{EOMchain} and \eqref{FMn}. 
Notice that $u_n \equiv \widetilde{u}_n/d$, $t \equiv \omega_1 \, \tilde{t}$, $\zeta \equiv \widetilde{\zeta}/\omega_1$, $\mu_n \equiv \widetilde \mu_n / d^3 \omega_1^2$, $r \equiv h/d$, $\Omega \equiv \omega_f/\omega_1$ and $F \equiv \widetilde{F} / d \, \omega_1^2$.
Hereafter, all parameters used in this work, including time, are non-dimensional.

Based on the above, the governing equations in their non-dimensional form are
\begin{equation}
	\label{EOMchain}
	\ddot{u}_n + 2\zeta \dot{u}_n + u_n + \Kappa \, \Delta^2 (u_n) + F_{M,n} = f_n \cos(\Omega t) , \quad 1 \le n \le N
\end{equation}
where $u_n$ is the normalized displacement of each unit, % and overdot denotes time derivative. %See \cite{JSV1} for a derivation of \eqref{EOMchain} and the corresponding physical system. 
$\zeta$ is normalized viscous damping coefficient for each unit, $\Kappa$ represents the normalized coupling force between adjacent units and $\Delta^2 (u_n) = 2u_n-u_{n+1}-u_{n-1}$ everywhere except at the boundaries, where $\Delta^2 (u_1) = u_1-u_2$ and $\Delta^2 (u_N) = u_N-u_{N-1}$ (free boundary conditions). %with $u_0=0=u_{N+1}$ (fixed boundary conditions). 
The normalized external force is given by $f_n = F$ for $n=1$ and zero otherwise. $F_{M,n}$ represents the only nonlinear force in the model and is given by
\begin{equation}
	\label{FMn}
	F_{M,n} = \mu_n \left( 
	  \frac{(1+u_n)}{((1+u_n)^2+r^2 )^{3/2}}
	 -\frac{(1-u_n)}{((1-u_n)^2+r^2 )^{3/2}}
	\right)
\end{equation}
where $r$ is a fixed parameter and $\mu_n$ is a control parameter that can be changed. %\revtext{Refer to~\ref{sec:repetition} for a physical interpretation of $r$.} 
In the case of an ordered periodic structure, we have $\mu_n=\mu_0$ for all $n$. For a disordered periodic structure, $\mu_n = \mu_0 + \delta \mu_n$ where $|\delta \mu_n|<|\mu_0|$. 
Expanding \eqref{FMn} about its trivial equilibrium point,  $u_n=0$, we obtain
\begin{equation}
	\label{Taylor}
	F_{M,n} \approx -\mu_n (a_1 u_n + a_3 u_n^3 + \cdots) \equiv k_1 u_n + k_3 u_n^3 + \cdots
\end{equation}
where $a_1$ and $a_3$ only depend on the (fixed) parameter $r$. 
The type and strength of nonlinearity can be tuned by changing the control parameter $\mu_n$. In particular, the strength of nonlinear terms increases when we increase the magnitude of $\mu_n$. Moreover, the sign of $\mu_n$ determines whether nonlinearity is of the hardening or softening type. When $\mu_n<0$, $k_3>0$ in \eqref{Taylor} and we have a hardening system. 

For small-amplitude vibrations, we can linearize \eqref{EOMchain} to get
\begin{equation}
	\label{EOMchainLin}
	\ddot{u}_n + 2\zeta \dot{u}_n + \omega_0^2 u_n + \Kappa \, \Delta^2 (u_n) = f_n \cos(\Omega t) 
\end{equation}
where 
\begin{equation}
	\label{omegaLin}
	\omega_0^2 = \omega_0^2(\mu_n) =  1 + k_1
\end{equation}
and $k_1$ is defined in \eqref{Taylor}.

We use a periodic structure with 10 units ($N=10$) throughout this work. We consider light damping, $\zeta = 0.005$, and weak coupling between adjacent units, $\Kappa = 0.05(1+k_1)$. The need for weak coupling  (strong response localization) will be motivated in Section~\ref{sec:disordered}. We set $r=0.1$. Unless otherwise noted, we use $\mu_0=-0.0270$, which gives $k_1=0.105$ and $k_3=0.2$ for the ordered system (hardening nonlinearity). This value of $\mu_0$ is chosen to ensure supratransmission occurs. The importance of the type of nonlinearity is addressed in Section~\ref{sec:supra}. The remaining two parameters are $F$ and $\Omega$, which are free parameters.

%%%%%%%%%%%%%%%%%%%%%%%%%%%%%%%%%%%%%%%%%%%%%%%%%%%%%%%%%%%%%%%%%%%%%%%%%%%%%%%%%%%%
\subsection{Exactly-Periodic Systems and Band Structure}
\label{sec:exactlyperiodic}

\begin{figure}[bht]
\centering
	\includegraphics[width=\linewidth]{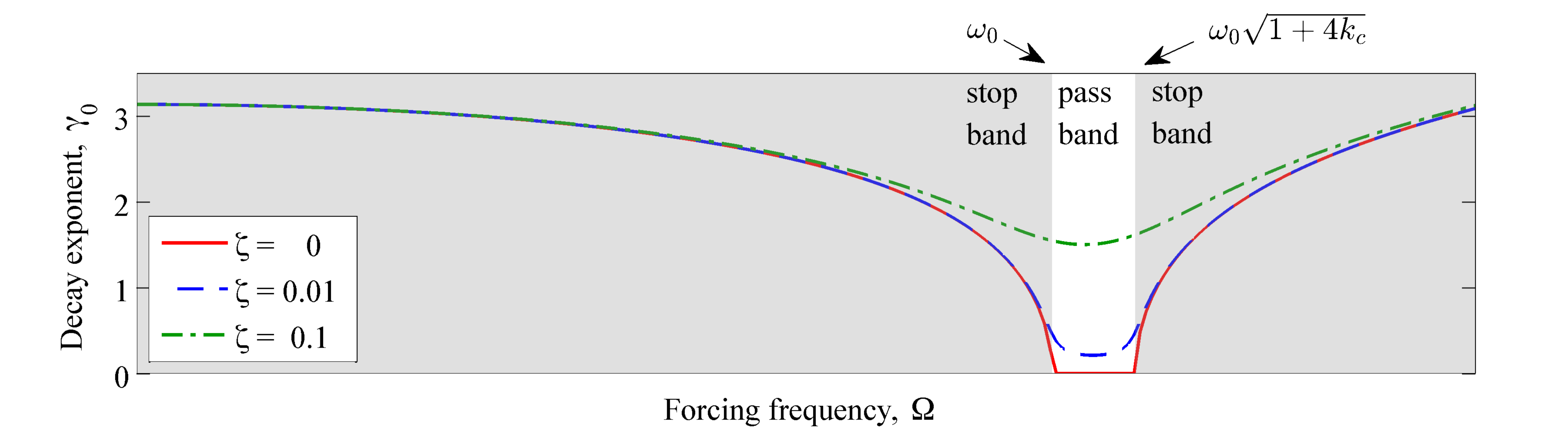}
	\caption{
	Decay exponent $\gamma_0$ for an infinitely-long exactly-periodic linear structure. The white and grey backgrounds indicate the pass and stop bands, respectively. When $\zeta>0$, the decay exponent is always positive, even within the pass band. 
	Anderson localization is relevant for frequencies within the pass band, whereas supratransmission occurs within the stop band 
	{\color{black}-- we will discuss these phenomena in subsequent sections. }
	}
	\label{fig:gamma0}
\end{figure}

 The solutions to the linear system \eqref{EOMchainLin} can be written as follows:
\begin{equation}
	\label{solutionLin}
	u_n(t) = u_1(t) e^{-\imath \qwave (n-1)} e^{-\gamma_0 (n-1)}
\end{equation}
where $\imath=\sqrt{-1}$, $\qwave$ is normalized wavenumber, $u_1(t)$ denotes the response of the first unit as a function of time, and $\gamma_0 \ge 0$ is a real-valued decay exponent. $u_1(t)=U_1 \exp(\imath \Omega t)$ in which $U_1$ is the complex-valued amplitude of motion. When $\gamma_0>0$, amplitudes of vibration attenuate exponentially through the chain and energy propagation is not possible over long distances. We get complete transmission when $\gamma_0=0$. Substituting  solution \eqref{solutionLin} into the governing equations \eqref{EOMchainLin}, we arrive at the following 
\begin{subequations}
	\label{bandLin}
	\begin{align}
		\label{bandLin1}
		\cosh \gamma_0 \cos \qwave &= 1 + (\omega_0^2 - \Omega^2)/(2 \Kappa) \\
		\label{bandLin2}
		\sinh \gamma_0 \sin \qwave &= (2\zeta \Omega)/(2 \Kappa)
	\end{align}
\end{subequations}
In the absence of damping ($\zeta=0$), we have $\gamma_0=0$ (complete transmission) for $\omega_0^2 \le \Omega \le \omega_0^2+4\Kappa$; this is the first pass band of the system. Outside this frequency range, $\gamma_0>0$ and waves will attenuate exponentially away from the source ($n=1$). 
This can be seen in Figure \ref{fig:gamma0} where $\gamma_0$ is plotted as a function of $\Omega$, for different values of damping. The area with a white background indicates frequencies for which $\gamma_0=0$ in the undamped system. We can see that the presence of damping results in spatial decay of response at all frequencies, even within the pass band. 
In particular, the influence of damping on the decay exponent is mostly significant within the pass band. Finally, we note that the analysis in this section is exact for an infinitely-long exactly-periodic linear structure. See \cite{husseinDamping} for more details on the influence of damping on the band structure of (ordered) linear periodic structures. 

%system('gs -o -q -sDEVICE=png256 -dEPSCrop -r300 -ojsv2_gamma0.png jsv2_gamma0.eps')

%%%%%%%%%%%%%%%%%%%%%%%%%%%%%%%%%%%%%%%%%%%%%%%%%%%%%%%%%%%%%%%%%%%%%%%%%%%%%%%%%%%%
%%%%%%%%%%%%%%%%%%%%%%%%%%%%%%%%%%%%%%%%%%%%%%%%%%%%%%%%%%%%%%%%%%%%%%%%%%%%%%%%%%%%
\subsection{Disorder-Borne Energy Localization within a Pass Band}
\label{sec:disordered}

Wave propagation within a linear periodic structure may be significantly influenced by presence of small disorder. In a disordered structure, waves scatter differently at the boundaries between adjacent units because the units are no longer identical. This normally results in spatial decay of response amplitude at frequencies within the pass band of the \revtextt{underlying} ordered structure. If we average the response over many different realizations of a prescribed disorder, then the response becomes localized to the source of excitation and decays exponentially away from it. This statistical phenomenon is known as \emph{Anderson localization} -- see \cite{anderson,Hodges82,anderson50,maynardPart2}. 

Disorder in engineering structures usually means small variations in structural parameters such as stiffness. These small irregularities can lead to significant \emph{qualitative} changes in the global dynamic response. This spatial confinement of energy, or localization, is called \emph{strong localization}~\cite{PierreWeakStrong}, and occurs when the strength of coupling between adjacent units is weak in comparison to the strength of disorder. 
In \emph{weak localization}, the coupling force is strong and damping effects dominate over disorder \cite{PierreWeakStrong,CastanierIndividual}. Even in undamped structures, weak localization effects are only significant at very long distances (at least a few hundred units~\cite{PierreWeakStrong}). 
Given that engineering structures always have damping and do not typically consist of such large number of units, only strong localization is relevant in the majority of engineering applications. Accordingly, we study a damped finite periodic structure with weak coupling. 
We consider a lightly damped structure and use a linear viscous damping model with a mass-proportional damping matrix. Refer to \cite{SPdamping1,SPdamping2} for a comprehensive review of general non-proportional and non-viscous damping models.

It is important to note that disorder-borne confinement of energy only occurs \emph{in an ensemble-average sense} and that individual realizations of disorder may behave very  differently. In particular, it is possible that the normal modes are localized away from the driving point for a particular realization. 
These anomalous realizations can have a significant influence on the average response of the ensemble, to the extent that using a linear average may not necessarily give the typical value (statistical \emph{mode}) of the ensemble \cite{HodgesEnsemble1,CastanierIndividual}. This is important in the case of weak localization \cite{chaAIAA}, where localization length scales are large. For a damped system, the contributions from anomalous realizations are much less significant because of the uniform decay caused by damping \cite{HodgesEnsemble1,CastanierIndividual}.

Anderson localization can occur as a result of disorder that is present in either the grounding springs (on-site potential), coupling springs (inter-site potential) or masses within the periodic structure \cite{kisselPhD}. It is argued \cite{PierreWeakStrong} that random masses and grounding springs influence the degree of localization in a similar manner, while random coupling springs have a weaker influence in comparison. Here, we only consider disorder that is applied to the linear grounding springs of the system. This type of disorder is sufficient to capture strong localization in our system.
In presence of such disorder, the linear governing equations of \eqref{EOMchainLin} are replaced with
\begin{equation}
	\label{EOMchainLinDis}
	\ddot{u}_n + 2\zeta \dot{u}_n + \omega_0^2 (1+\delta k_n ) u_n + \Kappa \, \Delta^2 (u_n) = f_n \cos(\Omega t) 
\end{equation}
where $\delta k_n$ are random numbers with a uniform probability density function. We assume that $\delta k_n$ are distributed independently around a zero mean, $<\delta k_n>=0$, such that $|\delta k_n| \le D$. We refer to $D$ as the \emph{strength of disorder}. 
We further introduce a parameter $C$ that denotes the \emph{strength of coupling} between units
\begin{equation}
	\label{coupling}
	C \equiv \Kappa / \omega_0^2
\end{equation}
Our earlier assumptions of weak coupling can now be expressed as $C<1$. The value of $C$ represents the ratio of coupling to grounding spring stiffness, and does not depend on the mass (coefficient of $\ddot u_n$). 

\begin{figure}[bth]
\centering
	\includegraphics[height=5cm]{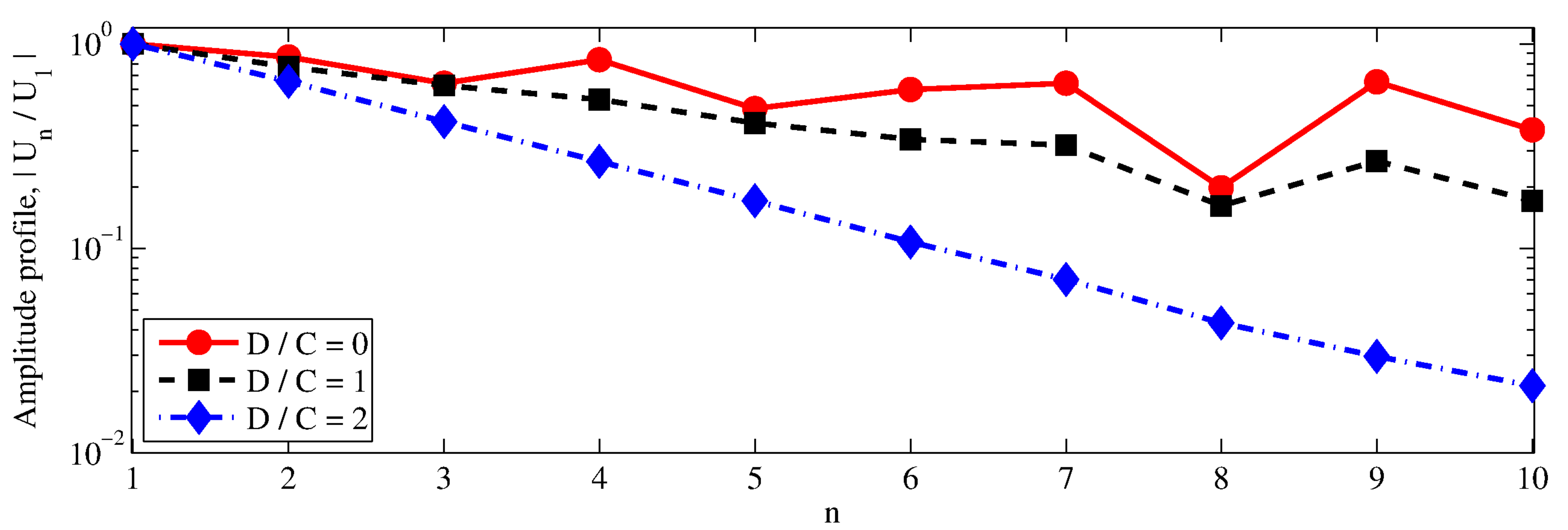}
	\caption{Influence of disorder on response localization at $\Omega = 1.12$, near the middle of the pass band. The average response becomes localized to the driven unit ($n=1$) as the value of $D/C$ increases. 
	The strength of coupling is kept constant, $C=0.05$. An ensemble of $10^5$ realizations are used for each non-zero value of $D$. Other system parameters are $N=10$, $\zeta=0.005$ and $\omega_0^2=1.05$. 
	%{\color{blue}complete caption}
	}
	\label{fig:anderson_Un}
\end{figure}

A crucial parameter that determines the degree of localization is $D/C$, the ratio of the strengths of disorder to coupling~\cite{Hodges82}. As this ratio increases, the degree of localization within the pass band increases as well. 
This is shown in Figure \ref{fig:anderson_Un} where we plot the normalized amplitude profile ($U_n/U_1$) for different strengths of disorder. The results for the disordered systems are obtained after averaging over an ensemble of $10^5$ realizations to ensure convergence.
We see in Figure \ref{fig:anderson_Un} that the response of the ordered structure ($D/C=0$) is extended through the system, and the small attenuation in response amplitude is due to damping. The oscillations in the response of the ordered system near the end of the structure are caused by the free boundary condition at $n=10$. 
Adding disorder results in smaller response amplitude in comparison with the ordered case. As the strength of disorder increases, the response becomes localized to $n=1$, where the external force is applied. 
%Also, notice that boundary effects at $n=10$ become insignificant for high values of $D/C$. This implies that disorder effects eventually subdue wave reflection at boundaries, rendering their contribution to the response unimportant.  
In addition, disorder effects eventually subdue wave reflections at the boundaries, which is why the boundary effects at $n=10$ become insignificant for high values of $D/C$.

%system('gs -o -q -sDEVICE=png256 -dEPSCrop -r300 -omcmat_anderson_Un.png mcmat_anderson_Un.eps')

\begin{figure}[bth]
\centering
	\includegraphics[width=\linewidth]{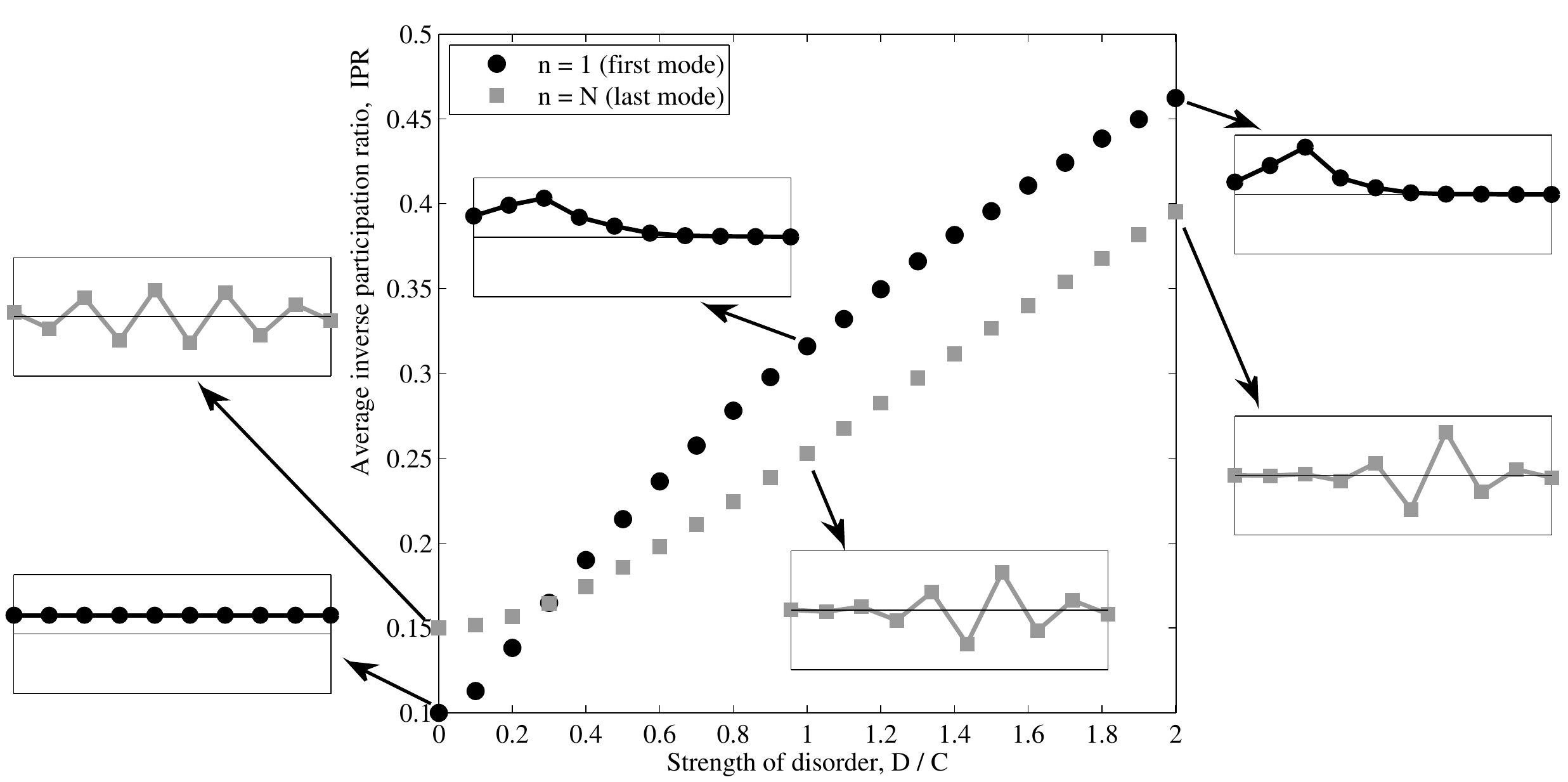}
	\caption{\revtext{
	Influence of disorder on spatial localization of the mode shapes of the structure. IPR, defined in~\eqref{ipr}, is plotted as a function of $D/C$ for the first and last mode shapes. The insets show the mode shape of a typical realization of disorder within the ensemble at $D/C=0,1,2$. The result for $n=1$ are shown in black and those for $n=N$ are shown in grey. 
	}}
	\label{fig:ipr}
\end{figure}

\revtext{
An important characteristic of disorder-borne localization is that the mode shapes of a disordered structure become spatially localized~\cite{Hodges82,iprRef}. This can be quantified using the inverse participation ratio (IPR), defined by
\begin{equation}
	\label{ipr}
	IPR = \frac{\sum_N^{n=1} (U_n^2)^2}{\left(\sum_N^{n=1} U_n^2\right)^2}
\end{equation}
where $\{U_n\}$ is the $n$-th mode shape of the system; e.g. see~\cite{iprRef} for more details. The IPR is a scalar with a value between $1/N$ and 1. If all the units are moving with the same amplitude (uniform response), then $\text{IPR}=1/N$. If only one unit is moving (absolute localization), then $\text{IPR}=1$. 
For the same ensemble used for Figure~\ref{fig:anderson_Un}, we have computed the average IPR for the first and last mode shapes. We show this in Figure~\ref{fig:ipr}, along with a typical mode shape of the ensemble at three values of $D/C$. We can see that the mode shapes become spatially localized as disorder becomes stronger. 
One can also use a modal expansion to express the response of the forced structure in terms of its mode shapes; in this light, the response localization we observed in Figure~\ref{fig:anderson_Un} can be explained based on spatial localization of the mode shapes~\cite{HodgesWoodASA}. 
}

%system('gs -o -q -sDEVICE=png256 -dEPSCrop -r300 -omcmat_anderson_Un.png mcmat_anderson_Un.eps')

%%%%%%%%%%%%%%%%%%%%%%%%%%%%%%%%%%%%%%%%%%%%%%%%%%%%%%%%%%%%%%%%%%%%%%%%%%%%%%%%%%%%
\subsection{Quantifying Response Localization}
\label{sec:exponent}

%Presenting the decay exponent (derivation in appendix). 

When there is exponential spatial decay of the response due to disorder, we expect 
\begin{equation}
	\label{xiMain}
	|U_n| \propto\exp(-\gamma_n n) %  e^{-\gamma n} %
\end{equation}
in an average sense. $\gamma_n$ is the localization factor or decay exponent and describes the average rate of exponential amplitude decay per unit. As the length of the periodic structure extends to infinity (equivalently, when averaged over many realizations), expression \eqref{xiMain} yields the correct decay rate \cite{HodgesWoodASA}. 
Calculation of the decay exponent merely based on \eqref{xiMain} involves solving for all response amplitudes and fitting an exponential curve to the data. 

There exist different approaches for studying the decay exponent analytically, such as using transfer-matrix~\cite{PierreWeakStrong} or receptance-matrix~\cite{MeadDisorder} formulations. In the former case, one could make use of the properties for products of random matrices~\cite{FurstenbergMatrix} and obtain asymptotic estimates for decay exponents~\cite{CastanierIndividual}. 
Alternatively, one could replace \eqref{xiMain} with 
\begin{equation}
	\label{xiHodges}
	|U_N| \equiv F \exp(-\gamma_N N)
\end{equation}
and use the properties of tridiaognal matrices to obtain expressions for $\gamma_N$ \cite{HerbertJones,HodgesWoodASA}. %This is the approach that we take in this work. 
It is important to note that \eqref{xiHodges} is true for any finite linear system and does not mean that the response is exponentially decaying. Nevertheless, if the response is exponentially decaying (as anticipated in Anderson localization), then the value of $\gamma_N$ obtained from averaging \eqref{xiHodges} represents the average decay rate in the limit of a very long structure (as $N\rightarrow\infty$). 

Based on the value of $\gamma_N$ alone one cannot draw any conclusions about the response of the periodic structure. In addition, because we are dealing with relatively short periodic structures in this work ($N=10$), the boundary effects will have an effect on $\gamma_N$. These make the interpretation of the decay factor, as defined in \eqref{xiHodges}, somewhat ambiguous. An alternative is to redefine the decay exponent by replacing $|U_N|$ in \eqref{xiHodges} with $|U_N/U_1|$; i.e. normalizing the response at the end of the chain with that of the driven unit.  Then \eqref{xiHodges} is replaced by
\begin{equation}
	\label{xiUNU1}
	\left | \frac{U_N}{U_1} \right | \equiv e^{-\gamma (N-1)} %|U_N/U_1| \equiv \exp\left(-\gamma (N-1) \right)%
\end{equation}
Notice that the multiplier for the decay exponent is now $(N-1)$ because a wave goes through $N-1$ units from $n=1$ to $n=N$. Based on definition \eqref{xiUNU1}, we obtain $\gamma > 0$ whenever the response has decayed through the structure; i.e. $|U_N| < |U_1|$. A negative value of the decay exponents is obtained whenever $|U_N| > |U_1|$; we have observed this only at resonance frequencies for structures with few number of units and very light damping.  
We emphasize that definition \eqref{xiUNU1} does not imply that the spatial decay envelop is exponential. 
See \ref{sec:xicompare} for a comparison of the three decay exponents in describing amplitude profiles. We also show for an ordered structure that $\gamma \rightarrow \gamma_0$ as $N \rightarrow \infty$. Accordingly, we use the decay exponent defined in \eqref{xiUNU1}.

%%%%%%%%%%%%%%%%%%%%%%%%%%%%%%%%%%%%%%%%%%%%%%%%%%%%%%%%%%%%%%%%%%%%%%%%%%%%%%%%%%%%
\subsection{Influence of Disorder on Band Structure}
\label{sec:anderson}

\begin{figure}[bht]
\centering
	\includegraphics[width=\linewidth]{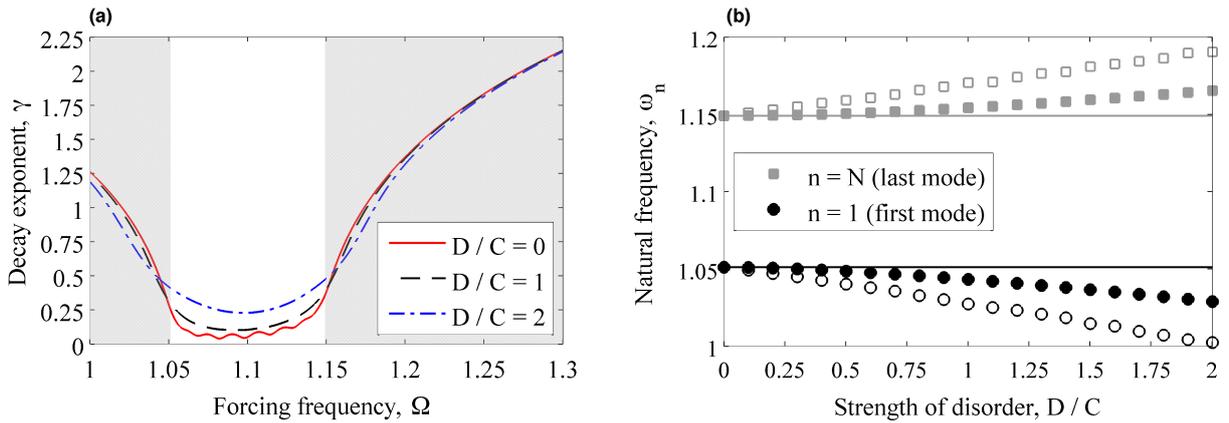}
	\caption{\revtext{
	Influence of disorder on the linear response of the structure. 
	(a) The average decay exponent $\gamma$, defined in \eqref{xiUNU1}, is plotted for different values of disorder. The grey area corresponds to the stop band. 
	(b) The average natural frequencies of the first ($\omega_1$) and last ($\omega_N$) modes are plotted as a function of $D/C$. 
	The black circles correspond to $\omega_1$ and grey squares to $\omega_N$. The empty markers correspond to the minimum and maximum values of each natural frequency within the ensemble. 
	The horizontal lines indicate the natural frequencies of the ordered structure. 
	}}
	\label{fig:omegan}
\end{figure}

The frequency of excitation plays a major role in determining the influence of disorder on the dynamic response of periodic structures. On average, less energy is transmitted to the end of a disordered structure at frequencies corresponding to the pass band of the corresponding ordered structure -- recall Figure \ref{fig:anderson_Un}. 
Figure~\ref{fig:omegan}(a) shows the average decay exponent $\gamma$ as a function of driving frequency $\Omega$ for different values of disorder. We can see that disorder results in increased values of decay exponent within the pass band of the ordered structure. Within the stop band, however, the decay exponent has a larger value for the ordered structure. Similar observations have been made in infinite undamped structures \cite{enhancedDisorder}. In addition, Figure~\ref{fig:omegan}(a) shows that disorder has the overall effect of slightly widening the pass band of a periodic structure. As a result, the transmitted energy is much smaller in a disordered structure but covers a wider frequency range in comparison to an ordered structure. %\textbf{[cite?]}
\revtext{The same conclusion can be made based on Figure~\ref{fig:omegan}(b), where we show the first and last natural frequencies of the structure as a function of the strength of disorder. We can see that as $D/C$ increases, $\omega_1$ decreases and $\omega_N$ increases on average.}

%system('gs -o -q -sDEVICE=png256 -dEPSCrop -r300 -omcmat_anderson_Un.png mcmat_anderson_Un.eps')

%%%%%%%%%%%%%%%%%%%%%%%%%%%%%%%%%%%%%%%%%%%%%%%%%%%%%%%%%%%%%%%%%%%%%%%%%%%%%%%%%%%%
%%%%%%%%%%%%%%%%%%%%%%%%%%%%%%%%%%%%%%%%%%%%%%%%%%%%%%%%%%%%%%%%%%%%%%%%%%%%%%%%%%%%
%%%%%%%%%%%%%%%%%%%%%%%%%%%%%%%%%%%%%%%%%%%%%%%%%%%%%%%%%%%%%%%%%%%%%%%%%%%%%%%%%%%%
%%%%%%%%%%%%%%%%%%%%%%%%%%%%%%%%%%%%%%%%%%%%%%%%%%%%%%%%%%%%%%%%%%%%%%%%%%%%%%%%%%%%
%%%%%%%%%%%%%%%%%%%%%%%%%%%%%%%%%%%%%%%%%%%%%%%%%%%%%%%%%%%%%%%%%%%%%%%%%%%%%%%%%%%%
%%%%%%%%%%%%%%%%%%%%%%%%%%%%%%%%%%%%%%%%%%%%%%%%%%%%%%%%%%%%%%%%%%%%%%%%%%%%%%%%%%%%
\section{Nonlinear Energy Transmission in Exactly-Periodic Structures}
\label{sec:nonlinearPeriodic}
%A brief introduction. Technical details such as analytical estimates and resonances with shifted pass bands will be provided when discussing the disordered case.

\subsection{Supratransmission: Energy Transmission from Excitation within a Stop Band} 
\label{sec:supra}

\begin{figure}[bht]
\centering
	\includegraphics[width=\linewidth]{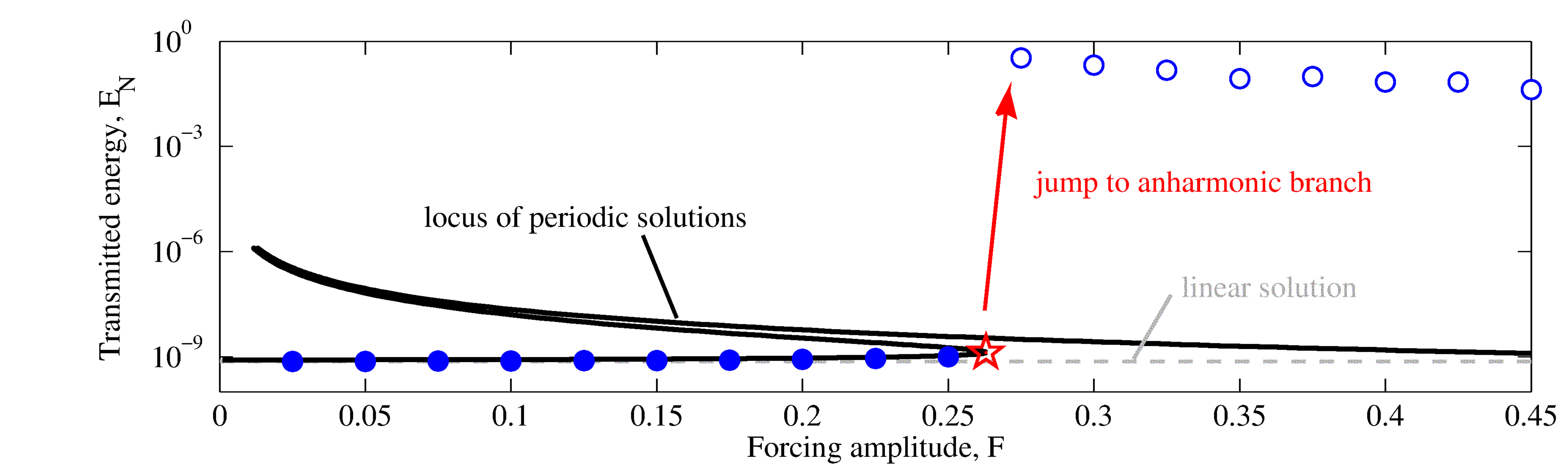}
	\caption{
	Energy transmitted to the end of the ordered structure ($E_N$) as a function of driving amplitude ($F$) at $\Omega=1.30$. The horizontal dashed grey line corresponds to the linear response of the system. The black curve  indicates the evolution of periodic solutions. 
	Circle markers indicates results from direct numerical integration (DNI) of the governing equations \eqref{EOMchain}, with filled markers indicating periodic response and empty markers indicating non-periodic response. 
	The solutions lose their linear stability through a saddle-node bifurcation, which is depicted by the red star. \revtext{The other periodic solution at $\Omega=1.30$ is unstable. Thus, the solution jumps to an anharmonic branch (empty circles)}.  
	}
	\label{fig:supra0}
\end{figure}

We consider the hardening system with $k_3=0.2$. We pick a forcing frequency above the pass band, $\Omega = 1.30$, for which energy transmission is prohibited in the linear operating range. If the driving amplitude $F$ is large enough, however, the periodic solutions lose their stability through a saddle-node bifurcation and jump to a different branch. 
This is shown in Figure \ref{fig:supra0}, where the average normalized energy at the end of the structure ($E_N$) is plotted as a function of driving amplitude. This time-averaged energy is defined at each unit as follows
\begin{equation}
	\label{En}
	E_n = \frac{1}{(m_2-m_1)T} \int_{m_1T}^{m_2T} \left(\frac{u_n(t)}{F}\right)^2 dt
\end{equation}
where $T=2\pi/\Omega$, $m_1=500$ and $m_2=2500$. In direct numerical integration (DNI), the equations of motion in \eqref{EOMchain} are solved using zero initial conditions. For each integration, the value of $F$ is increased gradually over the first 50 forcing cycles. $m_1$ is chosen such that the initial transients are passed. The results from this process are shown in Figure~\ref{fig:supra0} using the circle markers. 
\revtext{The locus of periodic solutions is also computed and shown in Figure~\ref{fig:supra0} using the solid curve. In this work, we use numerical continuation to trace the locus of periodic solutions. To do so, the governing equations are first recast as a boundary value problem and then continuation methods (as implemented in AUTO~\cite{doedel}) are used to follow these solution branches and their subsequent bifurcations; see \cite[Ch.~10]{kuznetsov} for details on continuation of periodic orbits.}
For periodic solutions, $E_n$ is evaluated over one forcing period ($m_2-m_1=1$). 
%The branch of periodic solutions is also computed using numerical continuation, as implemented in AUTO~\cite{doedel}. \revtext{In order to trace the locus of the periodic solutions and their subsequent bifurcations, the governing equations are first recast as a boundary value problem; e.g. see \cite[Ch.~10]{kuznetsov} for details on continuation of periodic orbits.} 

%system('gs -o -q -sDEVICE=png256 -dEPSCrop -r300 -ojsv2_supra.png jsv2_supra.eps')

\begin{figure}[bht]
\centering
	\includegraphics[width=\linewidth]{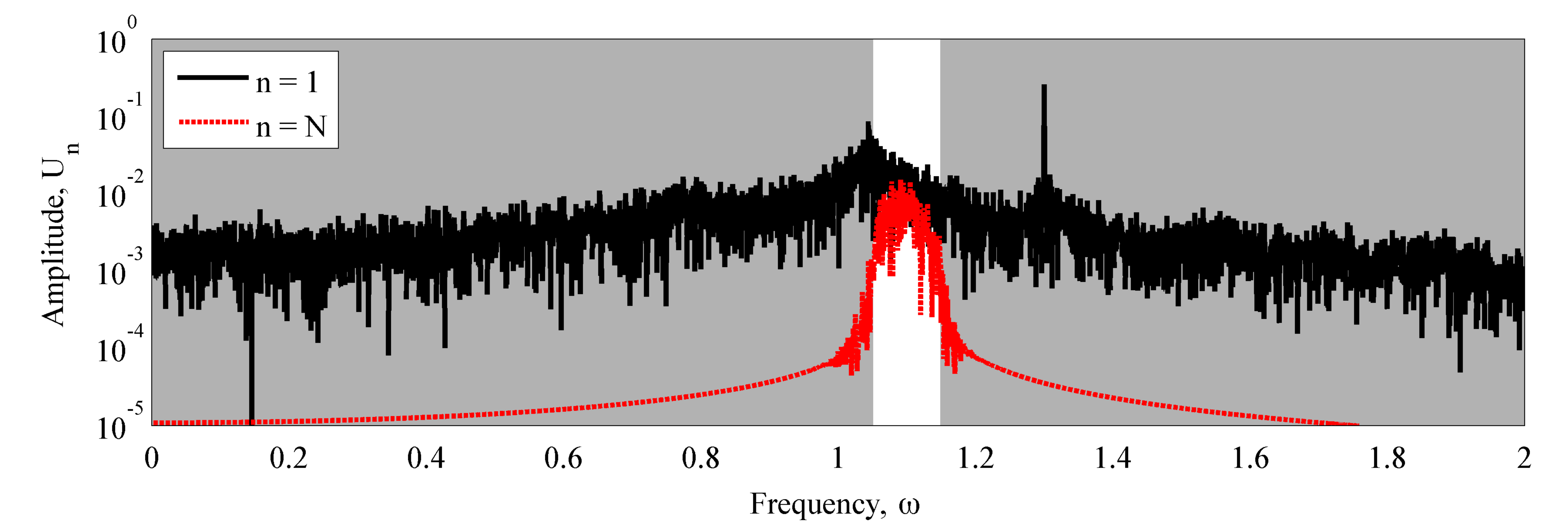}
	\caption{
	Frequency spectrum of the response for the ordered structure at the first ($n=1$) and last ($n=N=10$) units for $\Omega=1.30$ and $F=0.275$, just above the threshold in Figure \ref{fig:supra0}. 
	%The white background corresponds to the frequency range between the first and last linear natural frequencies of the ordered structure. 
	The grey area indicates the linear stop band.
	The peak at $\omega=1.30$ for $n=1$ corresponds to the driving frequency.
	}
	\label{fig:supra0spec}
\end{figure}
%system('gs -o -q -sDEVICE=png256 -dEPSCrop -r300 -ojsv2_supra_spec.png jsv2_supra_spec.eps')

\begin{figure}[hbt]
\centering
	\includegraphics[width=.98\linewidth]{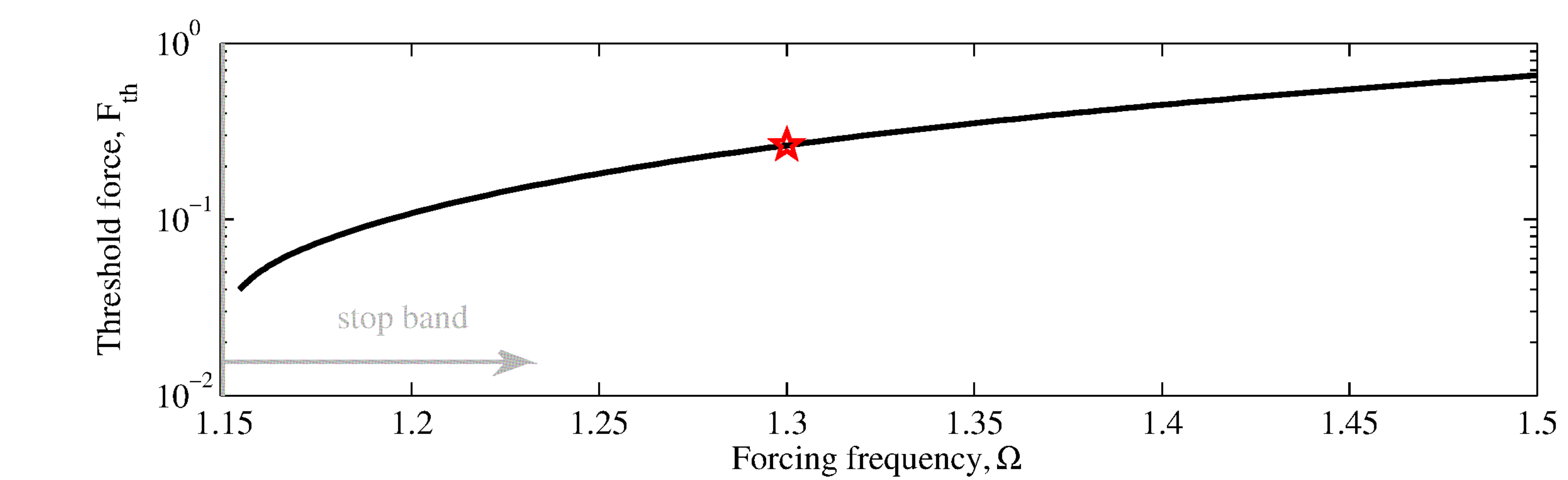}
	\caption{
	The threshold curve for the ordered structure, showing the dependence of the driving  amplitude at the onset of transmission, $F_{th}$, as a function of the driving frequency, $\Omega$. The star indicates $\Omega=1.30$, which was described in detail in Figures~\ref{fig:supra0}~and~\ref{fig:supra0spec}. \revtext{The threshold curve terminates near the pass band in a cusp.}
	%The grey area indicates the linear stop band. 
	%The white background corresponds to the frequency range between the first and last linear natural frequencies of the ordered structure. 
	}
	\label{fig:supra0FWcurve}
\end{figure}

We see in Figure~\ref{fig:supra0} that the response is periodic and follows the linear solution for low driving amplitudes. %At the saddle-node bifurcation, marked by a star, the response jumps to an anharmonic branch. This is accompanied by a large increase in the energy transmitted to the end of the structure. 
\revtext{At the saddle-node bifurcation, marked by a star, the periodic response becomes unstable. If another stable periodic branch exists at this point, the response typically jumps to that branch. A critical requirement then for supratransmission is that the upper harmonic branch is not stable at the saddle-node intersection. As a result, the solution jumps to a non-periodic branch. 
This jump is accompanied by a large increase in the energy transmitted to the end of the structure.  
If the upper periodic branch were stable, the increase in the transmitted energy would have been much smaller in comparison. This can be seen in Figure~\ref{fig:supra0}, by comparing the energies of the two periodic solutions near the onset of transmission, $F_{th} \approx 0.27$.
}

\revtext{An important characteristic of supratransmission can be observed in the frequency spectrum of the response above the transmission threshold}. Figure \ref{fig:supra0spec} shows the frequency spectrum for the first and last units of the periodic structure at $F=0.275$, just above the threshold. 
The frequency spectrum for each unit is obtained by taking the fast Fourier transform (FFT) of the corresponding time series after the initial transient has passed; no additional windowing or scaling is used.
We can see that the response at the first unit is \revtext{broadband and} highly nonlinear. As we go through the periodic structure to its other end, the frequency components within the linear stop band are highly attenuated \revtext{by dispersion}. Accordingly, the spectrum of the transmitted waves lies mainly within the pass band of the linear structure. \revtext{Thus supratransmission is a band-limited transmission phenomenon.}
The response of the system at driving amplitudes below the threshold is harmonic, with the frequency spectrum of each unit having a single dominant peak corresponding to the driving frequency (not shown). %, and their frequency spectra are not shown.

%It is emphasized that the mechanism responsible for this enhanced transmission is the saddle-node bifurcation \cite{supra_aubry,JSV1}. This is different from resonance of the driving force with the shifted pass band; see Section \ref{sec:resonance} for details. 

We can obtain the threshold force $F_{th}$ associated with different forcing frequencies $\Omega\,$, i.e. $F_{th}(\Omega)$, by tracing the location of the saddle-node bifurcation in Figure \ref{fig:supra0}. \revtext{We do this via numerical continuation.} We show the result of this computation in the $F-\Omega$ plane in Figure \ref{fig:supra0FWcurve}, and refer to it as the threshold curve of the system. 
%The star indicates $\Omega=1.30$, which we discussed in detail in Figures \ref{fig:supra0} and \ref{fig:supra0spec}. 
We see in Figure \ref{fig:supra0FWcurve} that the onset of transmission occurs at a higher forcing amplitude as we move away from the pass band. 
In a system with no damping, the threshold curve has a vertical asymptote at the closest linear natural frequency of the structure~\cite{supra_aubry}. 
For damped systems, on the other hand, the threshold curve could terminate (in a cusp) before reaching the pass band, as is the case in Figure \ref{fig:supra0FWcurve} -- see \cite{JSV1} for more details. 
%A full analysis of the influence of damping on the bifurcation structure within and near the pass band is outside the framework of this paper. 

\revtext{
The main influence of the type of nonlinearity (hardening or softening) is to determine on which side of the pass band supratransmission can occur: this is above the pass band in a hardening system and below it in a softening system~\cite{JSV1}.  Accordingly, we only consider driving frequencies on one side of the pass band for each type of nonlinearity. }

%%%%%%%%%%%%%%%%%%%%%%%%%%%%%%%%%%%%%%%%%%%%%%%%%%%%%%%%%%%%%%%%%%%%%%%%%%%%%%%%%%%%
\subsection{Enhanced Nonlinear Transmission from Excitation within a Pass Band} 
\label{sec:supraPassband}

\begin{figure}[hbt]
\centering
	\includegraphics[width=.9\linewidth]{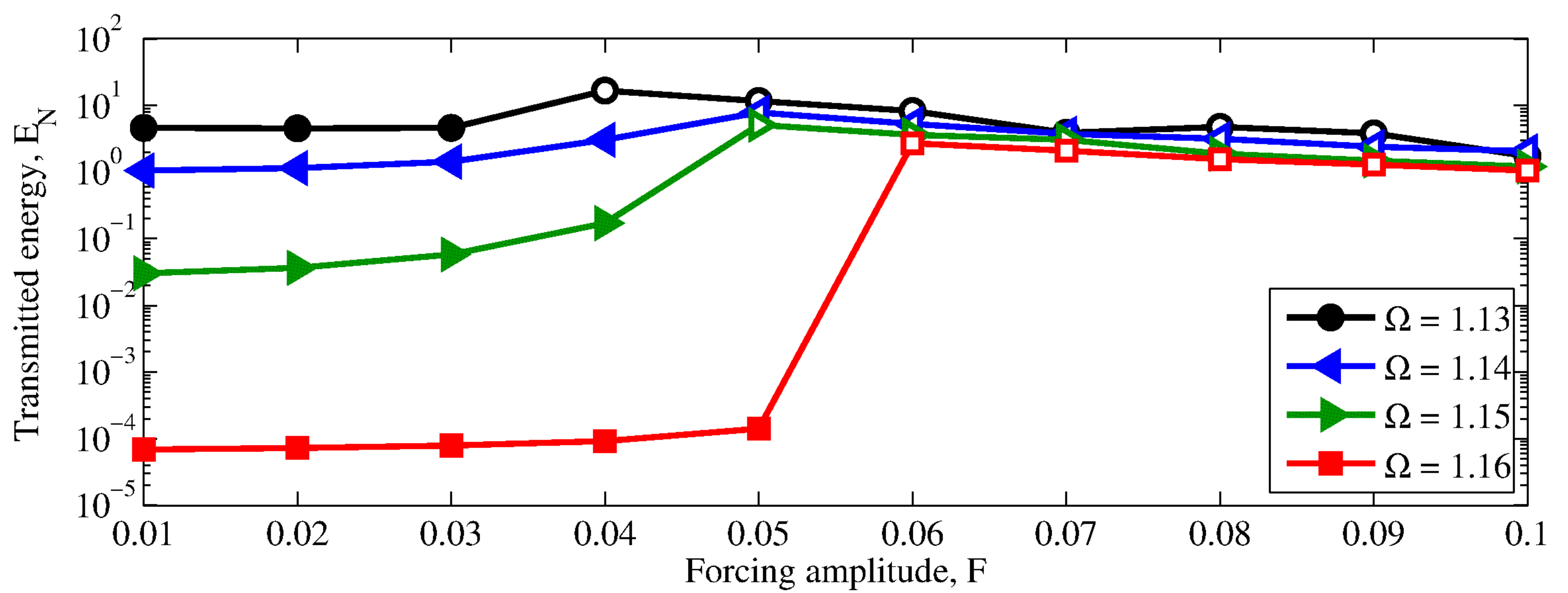}
	%{old_201509_withinPassband.png}%{old_mechanism_D_EN.png}
	\caption{
	Influence of forcing frequency ($\Omega$) on enhanced nonlinear energy transmission in a damped ordered structure. Energy at the end of the ordered structure ($E_N$) is plotted as a function of driving amplitude ($F$) for different forcing frequencies. The upper edge of the pass band is at $\omega_N\approx1.149$. $\Omega=1.13$ and $\Omega=1.14$ are inside the pass band, $\Omega=1.15$ is very close to the edge of the linear pass band, and $\Omega = 1.16$ is above the pass band. 
	Filled markers indicate periodic response and empty markers indicate non-periodic response. In all four cases, loss of stability leads to an increase in $E_N$. This increase is most significant when excitation is within the stop band. 
	}
	\label{fig:withinPassband}
\end{figure}

\revtext{The driving frequency plays a crucial role in determining energy transmission characteristics of a nonlinear periodic structure. To highlight this, we compare supratransmission (driving within a stop band) to the case in which an ordered structure is driven within its pass band.}
When the driving frequency lies within the pass band, energy transmission occurs in a linear fashion at low driving amplitudes, with a small decay occurring due to dissipative forces -- see Figures~\ref{fig:gamma0} and \ref{fig:omegan}(a). These linear solutions eventually lose their stability if the driving amplitude is increased. 
This loss of stability is not usually accompanied by a large increase in the transmitted energy, in strong contrast with what happens for stop band excitation. 
We see this in Figure~\ref{fig:withinPassband}, where we plot the transmitted energy as a function of $F$ at different forcing frequencies near the upper edge of the pass band. 

In all forcing frequencies shown in Figure~\ref{fig:withinPassband}, loss of stability is accompanied with an increase in the transmitted energy. 
At $\Omega = 1.16$, excitation is within the stop band and we see an increase in the transmitted energy over a few orders of magnitude. In comparison, the increase in transmitted energies is not as significant for driving frequencies within the pass band, namely at $\Omega = 1.13$ and $\Omega=1.14$.
$\Omega = 1.15$ is almost on the edge of the pass band and just before the beginning of the threshold curve. % (see  Figure~\ref{fig:supra0FWcurve}). 
Again, loss of stability leads to enhanced transmission, but not as significant as what happens at higher forcing frequencies within the stop band. 
\revtext{Within the pass band, the location of $\Omega$ with respect to the linear natural frequencies changes the onset of transmission ($F_{th}$), yet the qualitative behavior explained here remains the same; i.e. eventual loss of stability and an increase in $E_N$. If the structure were undamped, then the onset of transmission would have approached zero at the linear natural frequencies of the structure.}
%We do not further elaborate on this point.

Although loss of stability leads to enhanced transmission within both pass and stop bands, the increase in transmitted energies is much smaller within the pass band. 
One explanation is that the response is already extended throughout the system within the pass band, thus waves can reach the end of the structure without much attenuation -- compare the transmitted energies in Figure~\ref{fig:withinPassband} at low forcing amplitudes. Another factor could be the instability mechanism. Linear solutions lose their stability through saddle-node bifurcation at $\Omega = 1.16$ (and driving frequencies above it), whereas loss of stability occurs through a Neimark-Sacker bifurcation \cite[Ch.~5]{kuznetsov} at $\Omega=1.15$. The latter is also the typical instability mechanism inside the pass band. 
We have found damping to play a major role in determining \revtextt{the behavior of the Floquet multipliers and consequently} the mechanism leading to loss of stability in the vicinity of a pass band. %within a pass band. %
We discuss damping effects close to a pass band edge in Section~\ref{sec:supraDamping}. 
A full analysis of the influence of damping on the bifurcation structure inside the pass band lies outside the framework of this paper. 
\revtext{See~\cite{supra_disorder,supra_disorder_pikovsky} for detailed studies of disorder effects on transmission thresholds within a pass band.}

%Comment on bifurcations near and  within the pass band. Refer to Johansson \cite{supra_disorder} for further details. 

%%%%%%%%%%%%%%%%%%%%%%%%%%%%%%%%%%%%%%%%%%%%%%%%%%%%%%%%%%%%%%%%%%%%%%%%%%%%%%%%%%%%
\subsection{Influence of Damping on Supratransmission Near a Pass Band} 
\label{sec:supraDamping}

\begin{figure}[hbt]
\centering
	\includegraphics[width=\linewidth]{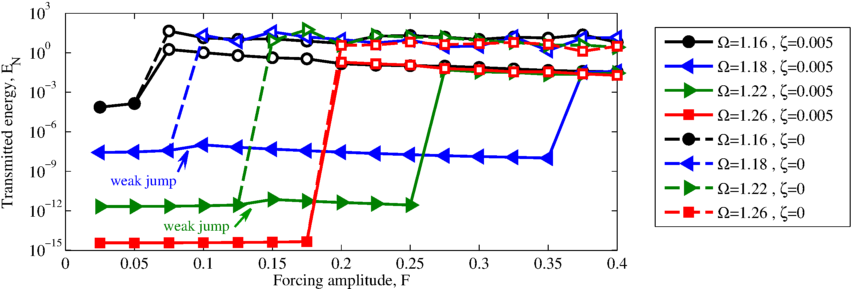}
	%{old_201509_withinPassband.png}%{old_mechanism_D_EN.png}
	\caption{\revtext{
	Influence of damping on supratransmission in the vicinity of a pass band for an ordered structure. Transmitted energy ($E_N$) is plotted as a function of driving amplitude ($F$) for different driving frequencies ($\Omega$). The upper edge of the pass band is at $\omega_N\approx1.149$. 
	Filled markers indicate periodic response and empty markers indicate non-periodic response. 
	Because damping is small, the threshold force ($F_{th}$) of the undamped structure is expected to be very close to that of the damped structure. 
	At $\Omega=1.16$ and $\Omega=1.26$, supratransmission occurs at the expected force threshold based on Figure~\ref{fig:supra0FWcurve}. 
	At $\Omega=1.18$ and $\Omega=1.22$, the response of the damped structure jumps to another stable periodic branch at the expected value of $F$ (indicated as `weak jump'). In contrast, supratransmission occurs for the undamped structure at the expected value of $F$. 
	}}
	\label{fig:nearPassband}
\end{figure}

\revtext{
In general, the force threshold at the onset of supratransmission ($F_{th}$) increases if the value of damping is increased~\cite{JSV1,supra_puri}. Close to the edge of a pass band, however, damping can play a more significant role. To illustrate this, we compare the transmitted energies in a damped ($\zeta=0.005$) and an undamped ($\zeta=0$) structure.
Figure~\ref{fig:nearPassband} shows $E_N$ as a function of $F$ at different values of $\Omega$ close to the pass band -- recall that the upper edge of the pass band is located at $\omega_N\approx1.149$. Given the low value of damping, we expect $F_{th}$ to be very similar for the two structures; thus, Figure~\ref{fig:supra0FWcurve} gives a good estimate for values of $F_{th}$ in the undamped structure.}

\revtext{
In Figure~\ref{fig:nearPassband}, we see that at $\Omega=1.16$ the onset of supratransmission occurs near $F_{th} \approx 0.05$ for both the damped and undamped structures. The transmitted energies above the threshold are lower for the damped structure, which can be easily attributed to energy loss through damping. 
At $\Omega=1.18$, we see that supratransmission occurs at the expected value of $F_{th} \approx 0.08$ for the undamped structure. In contrast, the damped structure has a relatively insignificant increase in $E_N$ between $F=0.075$ and $F=0.100$ (indicated as weak jump in Figure~\ref{fig:nearPassband}). The response of the damped structure is periodic below and above $F=0.08$, though with different amplitudes. Above $F=0.350$, supratransmission occurs in the damped structure as well. 
A similar difference between the damped and undamped structures is observed at $\Omega=1.22$. At $\Omega=1.26$ (and driving frequencies above it), supratransmission occurs around the same forcing thresholds for the damped and undamped structures, as initially expected.}

\begin{figure}[hbt]
\centering
	\includegraphics[width=.99\linewidth]{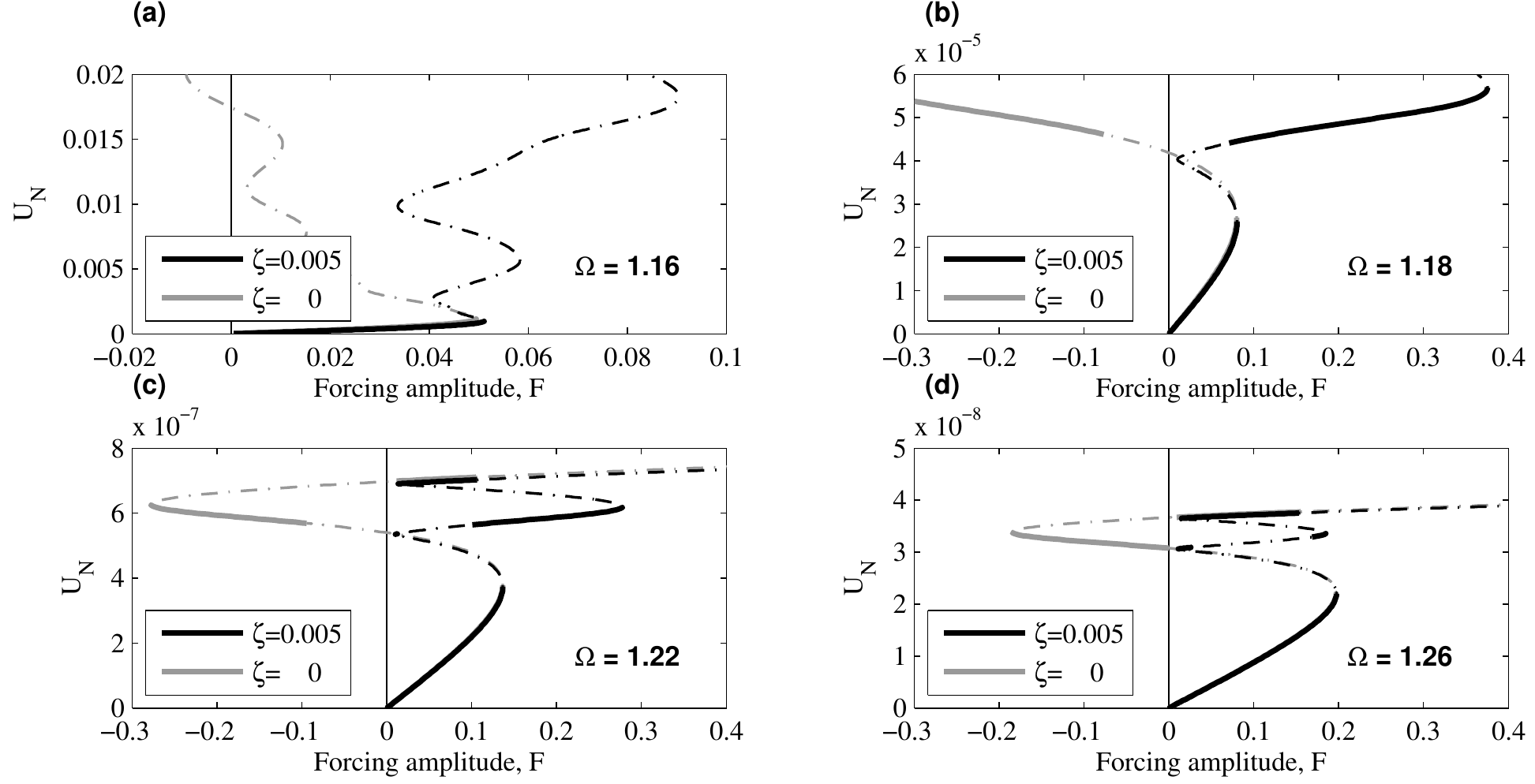}
	%{old_201509_withinPassband.png}%{old_mechanism_D_EN.png}
	\caption{\revtext{
	Influence of damping on the NLRM of the structure at different forcing frequencies: (a) $\Omega=1.16$, (b) $\Omega=1.18$, (c) $\Omega=1.22$, (d) $\Omega=1.26$. In each plot, the projection of the NLRM is plotted in the $U_N-F$ plane for $\zeta=0.005$ (black curve) and $\zeta=0$ (grey curve). For each NLRM, the thick solid sections represent stable solutions and thin dash-dotted portions represent unstable solutions. 
	}}
	\label{fig:NLRM}
\end{figure}

\revtext{
To understand the unexpected behavior of the damped structure when $1.18 \lesssim \Omega \lesssim 1.26$, we consider the evolution of the periodic solutions of the damped and undamped structures at different values of $\Omega$ as a function of $F$. This is referred to as the \emph{nonlinear response manifold} (NLRM)~\cite{supra_aubry,intrabandII}. Figure~\ref{fig:NLRM} shows the projection of the NLRMs on the $U_N-F$ plane for the same four values of $\Omega$ that are used in Figure~\ref{fig:nearPassband}. 
}

\revtext{
At $\Omega=1.16$, shown in Figure~\ref{fig:NLRM}(a), we see that the two NLRMs start from the origin and follow the linear solution ({i.e.} $U_N \propto F$) for small values of $F$. At the first turning point (TP1), the periodic solutions lose their stability through a saddle-node bifurcation. Because neighboring periodic solutions either do not exist or are unstable at this point, the solution jumps to a non-periodic branch. This is accompanied by the large increase in transmitted energies shown in Figure~\ref{fig:nearPassband}. 
After TP1, the undamped response normally crosses the zero-force axis ($F=0$) multiple times. These zero-crossings correspond to non-zero time-periodic solutions in the undamped system that are spatially localized to the driven unit -- these solutions are called discrete breathers~(DB). See~\cite{supra_aubry} for more details about the zero-crossings and the associated DBs. For a damped structure, it is important to note that the NLRM does not cross the zero-force axis because the structure can no longer sustain steady-state motion with non-zero amplitude; see also~\cite{maniadisDamping}. 
}

%This is defined as the manifold of initial conditions with zero initial velocities that give a time-reversible periodic solution at a given driving frequency. At the first turning point of the NLRM for a non-dissipative system, the stable periodic solution that continues from the linear response collides with another unstable periodic solution that continues from a discrete breather at zero driving amplitude (i.e. a saddle-node bifurcation).

\revtext{
At $\Omega=1.18$, shown in Figure~\ref{fig:NLRM}(b), we see that another stable periodic solution exists at TP1 for the damped structure. As a result, the solution jumps to that solution branch when $F$ is increased beyond its value at TP1. The response at this upper periodic branch is dominantly harmonic (with frequency $\Omega$) and has a higher value of $E_N$ than the linear solution -- see Figure~\ref{fig:nearPassband}. Nevertheless, supratransmission does not occur until the third turning point (TP3) of the NLRM around $F\approx 0.375$. For the undamped structure, other solutions at TP1 are unstable and the solution jumps to a non-periodic branch. 
}

\revtext{
At $\Omega=1.22$, shown in Figure~\ref{fig:NLRM}(c), the situation is similar to what happens at $\Omega=1.18$. The main difference is the range over which the upper periodic branch is stable: compared to $\Omega=1.18$, TP3 occurs at a lower value of $F$. 
Although the undamped NLRM has stable portions, we have not observed any situation in which the stable branch extends to TP1 -- the same observation is made in~\cite{supra_aubry}. %, though this has not been proved to be always true. 
Thus, supratransmission occurs at TP1 for the undamped structure. 
By $\Omega=1.26$, shown in Figure~\ref{fig:NLRM}(d), the damped NLRM has changed such that TP3 occurs again at a lower value of $F$ than TP1. Accordingly, supratransmission occurs at TP1 for both the damped and undamped structures. 
}

\revtext{
Figure~\ref{fig:NLRM} clearly shows that the stability of the upper branch of NLRM at the TP1 depends on the existence of damping. We have found that disorder can also play a role here. 
Investigating the necessary/sufficient conditions for this to occur sets forth a very interesting problem. Knowing the location of TP3 or subsequent turning points would not suffice for this purpose because the upper branch can change stability between turning points~(via Neimark-Sacker bifurcation), as seen in Figures~\ref{fig:NLRM}(b-d). A systematic investigation of the influence of damping and disorder on the stability of the upper branch at TP1 can be done using numerical continuation. This study, however, falls outside the scope of our present work {-- it is also of less relevance in applications based on supratransmission because supratransmission is significant at frequencies away from a pass band.}
Consequently, we will only consider driving frequencies for which supratransmission occurs via jump to an anharmonic branch at the first turning point of the nonlinear response manifold~\cite{supra_aubry,supra_disorder}. In this light, we will consider $\Omega \ge 1.25$ in Section~\ref{sec:supraDisordered}. 
}

\revtext{
We note that in addition to the effects discussed above, increasing damping may eventually eliminate supratransmission. This happens because damping can modify the topology of the NLRM such that it no longer has a saddle-node bifurcation. 
We do not further elaborate on this point because it has already been discussed in~\cite{JSV1}.
}

%%%%%%%%%%%%%%%%%%%%%%%%%%%%%%%%%%%%%%%%%%%%%%%%%%%%%%%%%%%%%%%%%%%%%%%%%%%%%%%%%%%%
%%%%%%%%%%%%%%%%%%%%%%%%%%%%%%%%%%%%%%%%%%%%%%%%%%%%%%%%%%%%%%%%%%%%%%%%%%%%%%%%%%%%
%%%%%%%%%%%%%%%%%%%%%%%%%%%%%%%%%%%%%%%%%%%%%%%%%%%%%%%%%%%%%%%%%%%%%%%%%%%%%%%%%%%%
%%%%%%%%%%%%%%%%%%%%%%%%%%%%%%%%%%%%%%%%%%%%%%%%%%%%%%%%%%%%%%%%%%%%%%%%%%%%%%%%%%%%
%%%%%%%%%%%%%%%%%%%%%%%%%%%%%%%%%%%%%%%%%%%%%%%%%%%%%%%%%%%%%%%%%%%%%%%%%%%%%%%%%%%%
%%%%%%%%%%%%%%%%%%%%%%%%%%%%%%%%%%%%%%%%%%%%%%%%%%%%%%%%%%%%%%%%%%%%%%%%%%%%%%%%%%%%
%%%%%%%%%%%%%%%%%%%%%%%%%%%%%%%%%%%%%%%%%%%%%%%%%%%%%%%%%%%%%%%%%%%%%%%%%%%%%%%%%%%%
\section{Transmission Thresholds in Disordered Nonlinear Periodic Structures}
\label{sec:supraDisordered}

We investigate the influence of disorder on supratransmission in this section. Similar to Section \ref{sec:disordered}, we consider disorder as random linear spring constants taken from a uniform distribution, with $-D \le \delta k_n \le D$. 
%We use the same type of disorder that was used in Section \ref{sec:disordered}, which is including linear additive irregularities in grounding springs; i.e. $\delta k_n$ in \eqref{EOMchainLinDis}. 
We consider two strengths of disorder, $D/C=1$ and $D/C=2$. Notice that $C=0.05$ is kept constant and the strength of disorder is varied by changing $D$. 
With the exception of Section \ref{sec:examples}, the results presented here are pertinent in an ensemble-average sense, meaning that they describe the behavior of a typical disordered structure; i.e. the statistical mode of the ensemble. %Individual realizations of a disordered structure may behave differently. 

\revtext{
Throughout this section, we study the hardening system with $k_3=0.2$ (similar to Section 3). Accordingly, we only consider forcing frequencies above the pass band because supratransmission occurs above the pass band in a hardening system~\cite{JSV1}. 
The upper edge of the pass band is located at $\omega_N \approx 1.149$ in this case. 
As explained in Section~\ref{sec:supraDamping}, we only consider $\Omega \ge 1.25$ to focus on supratransmission occurring at the first turning point of the nonlinear response manifold. 
%Based on the discussion in Section~\ref{sec:supraDamping}, we only consider $\Omega \ge 1.25$ to avoid the non.
Qualitatively similar results are expected if a softening nonlinearity is chosen and forcing frequencies are below the pass band; see~\cite{ISOT}. Both hardening and softening types of nonlinearity will be considered in Section~\ref{sec:evaluation}. 
}

\subsection{Loss of Stability Leads to Enhanced Transmission}
\label{sec:examples}

\begin{figure}[hbt]
\centering
	\includegraphics[width=\linewidth]{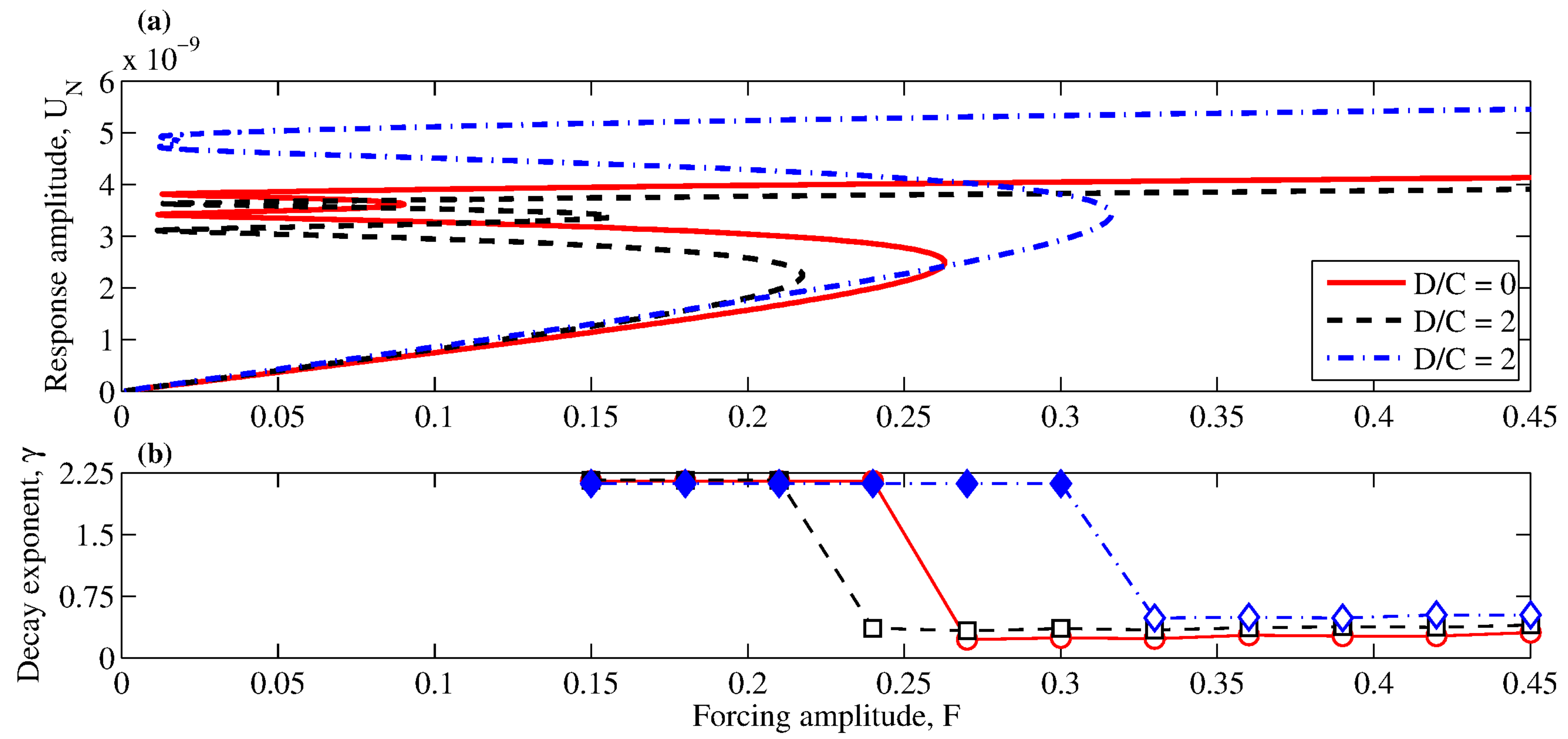}
	\caption{Influence of disorder on the onset of transmission at $\Omega=1.30$. (a) the locus of periodic solutions in the $U_N-F$ plane, the first turning point corresponding to loss of stability; (b) decay exponent as a function of $F$ obtained from direct numerical integration (DNI) of the governing equations. 
	Solid red curve corresponds to the periodic structure ($D=0$), black dashed curve and blue dash-dotted curve correspond to two different realizations of the disordered structure, both with $D/C=2$. 
	For both realizations, $\sum \delta k _n \approx 0$. Apart from the presence of disorder, all other system parameters are the same as in Section \ref{sec:nonlinearPeriodic}. 
	The data depicted here by red solid curves corresponds to the same configuration as in Figure \ref{fig:supra0}. The two plots (a) and (b) have the same horizontal axis.}
	\label{fig:examples}
\end{figure}

%The enhanced transmission within a stop band persists in presence of disorder.
For a given forcing frequency, the onset of transmission in a disordered structure may occur at a different value of the driving amplitude when compared with the corresponding ordered structure. The change in threshold force amplitude, and whether it occurs or not, depends on the particular realization that is being considered. We have shown this in Figure \ref{fig:examples} for two different realizations of disorder with $D/C=2$.

%system('gs -o -q -sDEVICE=png256 -dEPSCrop -r300 -omcmat_anderson_Un.png mcmat_anderson_Un.eps')

Figure \ref{fig:examples}(a) shows the evolution of the periodic solutions as a function of the driving amplitude, $F$. For low driving amplitudes (near the origin), the response amplitude increases linearly with $F$, as expected from linear theory. As $F$ increases, there is a turning point in the response~(saddle-node bifurcation); cf.~Figure~\ref{fig:supra0}. 
The solution then jumps to a non-periodic branch. Depending on the realization being considered, the onset of transmission may be lower or higher than the onset for the ordered structure.  Results from direct numerical integration  of \eqref{EOMchain} are shown in Figure \ref{fig:examples}(b). We can see that there is a significant decrease in the decay exponent that occurs at a value of $F$ consistent with the turning points in Figure \ref{fig:examples}(a).

% %%%%%%%%%%%%%%%%%%%%%%%%%%%%%%%%%%%%%%%%%%%%%%%%%%%%%%%%%%%%%%%%%%%%%%%%%%%%%%%%%%%%
% \subsection{Resonances with the Shifted Pass Band} 
% \label{sec:resonance}
% 
% Compare the onset of transmission with the shifted pass band. 

%%%%%%%%%%%%%%%%%%%%%%%%%%%%%%%%%%%%%%%%%%%%%%%%%%%%%%%%%%%%%%%%%%%%%%%%%%%%%%%%%%%%
%%%%%%%%%%%%%%%%%%%%%%%%%%%%%%%%%%%%%%%%%%%%%%%%%%%%%%%%%%%%%%%%%%%%%%%%%%%%%%%%%%%%
%%%%%%%%%%%%%%%%%%%%%%%%%%%%%%%%%%%%%%%%%%%%%%%%%%%%%%%%%%%%%%%%%%%%%%%%%%%%%%%%%%%%
%%%%%%%%%%%%%%%%%%%%%%%%%%%%%%%%%%%%%%%%%%%%%%%%%%%%%%%%%%%%%%%%%%%%%%%%%%%%%%%%%%%%
%%%%%%%%%%%%%%%%%%%%%%%%%%%%%%%%%%%%%%%%%%%%%%%%%%%%%%%%%%%%%%%%%%%%%%%%%%%%%%%%%%%%
% \section{Overall Influence of Disorder on Transmission Thresholds}
% \label{sec:supraStats}
% 
% 
% In this section, we discuss the statistical effects of disorder on supratransmission, when results are averaged over many realizations within an ensemble. We use the same type of disorder that was used in Section \ref{sec:disordered}. 

%%%%%%%%%%%%%%%%%%%%%%%%%%%%%%%%%%%%%%%%%%%%%%%%%%%%%%%%%%%%%%%%%%%%%%%%%%%%%%%%%%%%
%%%%%%%%%%%%%%%%%%%%%%%%%%%%%%%%%%%%%%%%%%%%%%%%%%%%%%%%%%%%%%%%%%%%%%%%%%%%%%%%%%%%
\subsection{Onset of Transmission Remains Unchanged on Average}
\label{sec:averageFW}

\begin{figure}[hbt]
\centering
	\includegraphics[width=\linewidth]{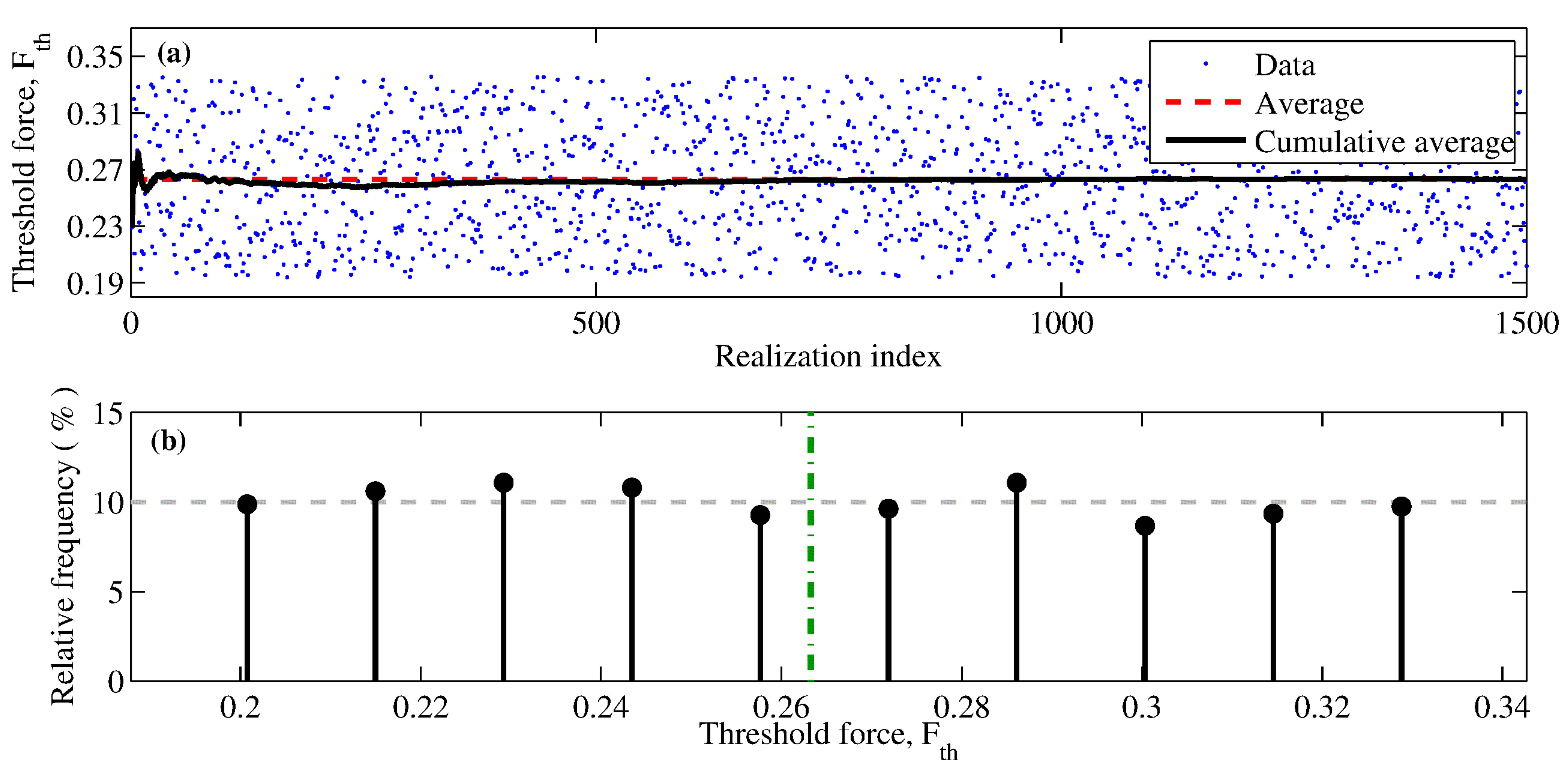}
	\caption{Influence of disorder on the onset of transmission $F_{th}$ at $\Omega=1.30$ for an ensemble of 1500 realizations with $D/C=2$. The value of $F_{th}$ for each realization is obtained by numerical continuation. 
	(a) Individual values of $F_{th}$ for each realization are shown by dots, the cumulative average is shown by the black solid curve and the total average is shown by the red dashed line. 
	(b) The relative frequency of occurrence of $F_{th}$ is shown by solid back lines. The value of $F_{th}$ for the ordered system is shown using the vertical dash-dotted line. The horizontal dashed line indicates relative frequency of 10\%.  
	The results suggest that on average the onset of transmission is the same for ordered and disordered system.  
}
	\label{fig:statsFth}
\end{figure}

\begin{figure}[hbt]
\centering
	\includegraphics[width=.9\linewidth]{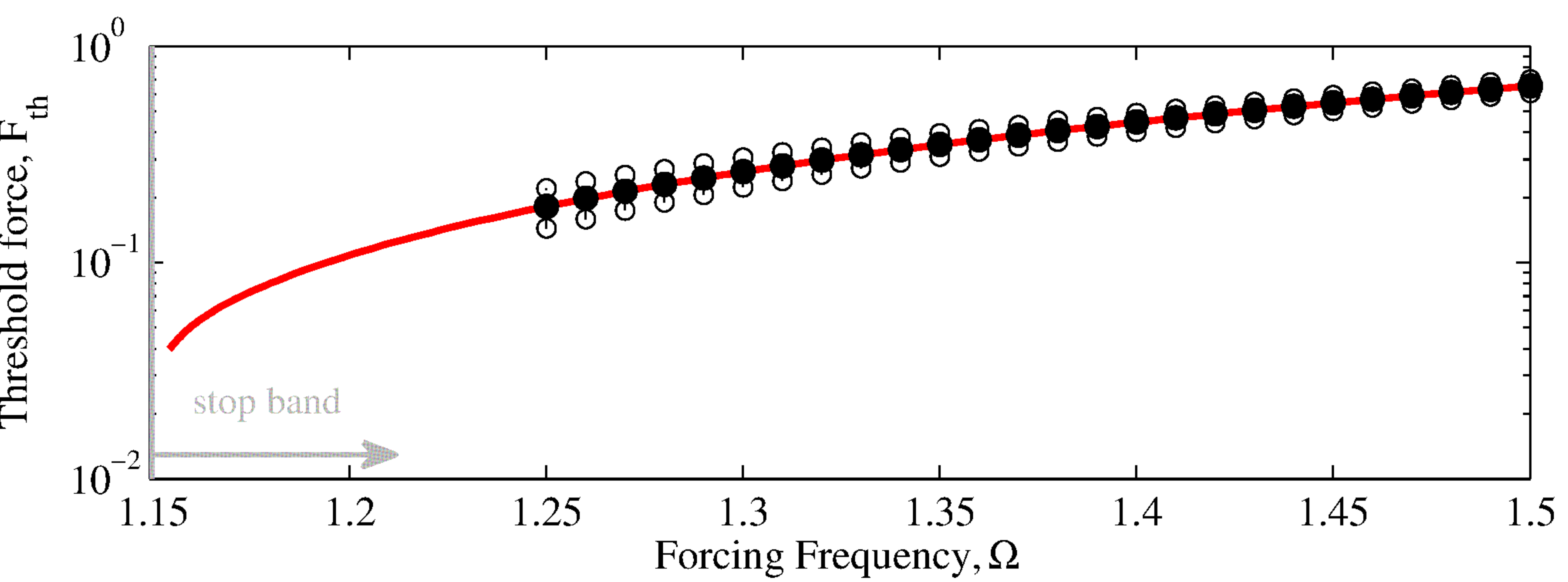}
	\caption{
	Influence of linear disorder on the threshold force at different forcing frequencies for the hardening structure. The red solid curve shows the threshold curve for an ordered structure (cf. Figure \ref{fig:supra0FWcurve}). 
	At each forcing frequency, the filled circles show the average value of $F_{th}$ and the empty circles show the corresponding value of standard deviation. We see that the average threshold curve for a disordered system is the same as the threshold curve for an ordered system. The standard deviation of $F_{th}$ decreases as we move away from the pass band. 
	}
	\label{fig:statsFW}
\end{figure}

Knowing that the onset of transmission occurs at different forcing amplitudes depending on the specific realization of a given disorder, we want to know the average value of the onset of transmission in an ensemble-average sense. 
To find the answer, we have computed the exact numerical value of the threshold force $F_{th}(\Omega)$ using numerical continuation. 
Figure \ref{fig:statsFth}(a) shows the values of $F_{th}$ at $\Omega=1.30$ for an ensemble of 1500 realizations with $D/C=2$. Comparing the cumulative average with the total average indicates that the results have converged. The relative frequency of occurrence of $F_{th}$ within the same ensemble is shown in Figure \ref{fig:statsFth}(b). 
The average value of $F_{th}$ for the disordered system was found to be the same as that of the ordered system. Moreover, we can see in Figure \ref{fig:statsFth}(b) that individual values of $F_{th}$ for the disordered system are spread uniformly around the onset of transmission for the ordered system. We have made similar observations at other values of $\Omega$ as well (not shown here).

Figure \ref{fig:statsFW} shows the average values of $F_{th}$ and their standard deviations  for $D/C=2$ at different values of $\Omega$ away from the linear pass band. 
The average values show that, \emph{in an ensemble-average sense}, the onset of supratransmission is the same in ordered and disordered systems. 
For a fixed strength of disorder, we observe that the standard deviation of $F_{th}$ decreases as $\Omega$ moves away from the pass band. This implies that as $\Omega$ moves farther into the stop band, dispersion (due to Bragg scattering) is more dominant than disorder effects. This is in fact similar to how disorder affects the linear response of the system. As we showed in Section \ref{sec:anderson}, the most significant influence of disorder occurs at frequencies within and near the pass band of the system. We found similar results for $D/C=1$. 

%Returning to Figure \ref{fig:histogram}(a), we notice that the influence of disorder on the transmitted energies is different below and above the threshold. 

%We recall from Section \ref{sec:anderson} that disorder results in slight increase of linear energy transmission within the stop band. This increased transmission is insignificant except in close vicinity of the pass band, and is negligible in comparison with the enhanced transmission due to instability (supratransmission). Comparing the decay exponents at $\Omega=1.30$ in Figures \ref{fig:anderson_exponent} and \ref{fig:examples}(b) clarifies this point.

%%%%%%%%%%%%%%%%%%%%%%%%%%%%%%%%%%%%%%%%%%%%%%%%%%%%%%%%%%%%%%%%%%%%%%%%%%%%%%%%%%%%
%%%%%%%%%%%%%%%%%%%%%%%%%%%%%%%%%%%%%%%%%%%%%%%%%%%%%%%%%%%%%%%%%%%%%%%%%%%%%%%%%%%%
\subsection{Energy Profiles of Transmitted Waves}
\label{sec:Eprofiles}

\begin{figure}[hbt]
\centering
	\includegraphics[width=\linewidth]{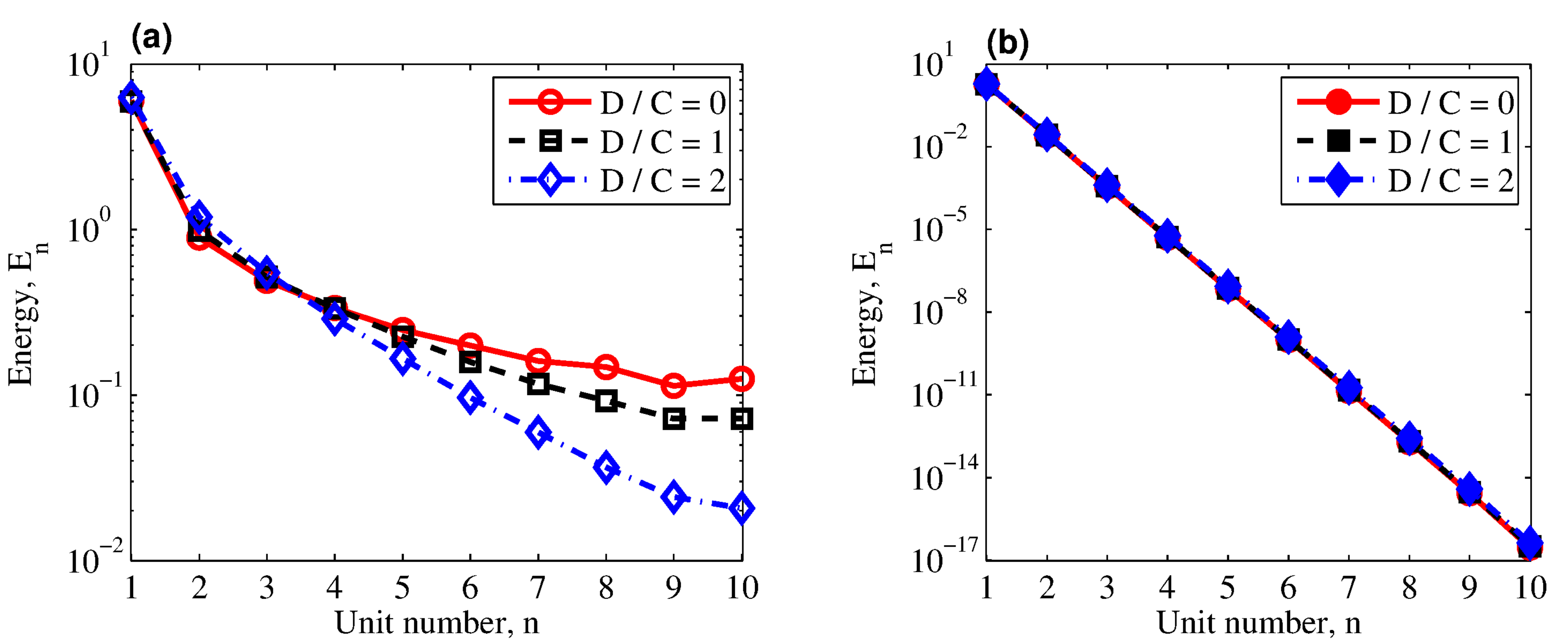}
	\caption{
	Average energy profiles at $\Omega = 1.30$ for (a) the nonlinear system above the onset of transmission, (b) the linear system. Notice the difference between the vertical axes in (a) and (b). Above the threshold, disorder results in localization of energy to the driven unit. 
	}
	\label{fig:Eprofiles}
\end{figure}

%Because we are within a stop band, the response amplitude decays exponentially as long as we are below the threshold. Here, we explore how disorder changes the average response above the threshold. We consider two strengths of disorder, $D/C=1$ and $D/C=2$. 
We explore how disorder changes the average response above the transmission threshold for two strengths of disorder, $D/C=1$ and $D/C=2$. 
For each disorder strength, we consider an ensemble of 1500 disordered structures; this ensemble is large enough that energies converge to their average values. For every realization within an ensemble, we compute the energy for each unit ($E_n$) at a forcing amplitude 5\% above the onset of transmission -- $E_n$ is defined in \eqref{En}. Average energies at each strength of disorder are then obtained by averaging energies over the entire ensemble. The results presented here are for $\Omega = 1.30$. 

Figure \ref{fig:Eprofiles}(a) shows the average energy profiles for the nonlinear system above the onset of transmission. We see that energies decay exponentially through the structure, particularly between $n=2$ and $n=N-1$ (away from the boundaries). 
Moreover, energy becomes more localized to the driven unit as the strength of disorder is increased. 
These energy profiles are drastically different from the energy profiles below the threshold, where the response is very similar to the linear response shown in Figure \ref{fig:Eprofiles}(b) -- given that the response below the threshold is very similar to the linear response, we have used the linear energy profiles in Figure \ref{fig:Eprofiles}(b). 
%Comparing Figures \ref{fig:Eprofiles}(a) and \ref{fig:Eprofiles}(b), we notice that disorder has a more significant influence on the response of the structure above the transmission threshold than below it; we will explain this in Section \ref{sec:spectra}. 
%We also note that disorder has a more significant influence on the response of the structure above the threshold than below it. This is because we are driving the structure within its stop band. While disorder has a very insignificant influence on the linear dynamics, it hinders energy transmission above the threshold in a nonlinear system. 
%We explain why disorder decreases the transmitted energies by studying the average frequency spectra of the transmitted waves in Section \ref{sec:spectra}. 
{We notice in Figure~\ref{fig:Eprofiles} that disorder effects are more significant above the transmission threshold (Figure~\ref{fig:Eprofiles}(a)) than below it (Figure~\ref{fig:Eprofiles}(b)). We explain this in more detail in the following section.} %Section \ref{sec:spectra}. 

%%%%%%%%%%%%%%%%%%%%%%%%%%%%%%%%%%%%%%%%%%%%%%%%%%%%%%%%%%%%%%%%%%%%%%%%%%%%%%%%%%%%
%%%%%%%%%%%%%%%%%%%%%%%%%%%%%%%%%%%%%%%%%%%%%%%%%%%%%%%%%%%%%%%%%%%%%%%%%%%%%%%%%%%%
\subsection{Average Frequency Spectra Above the Threshold}
\label{sec:spectra}

\begin{figure}[hbt]
\centering
	\includegraphics[width=\linewidth]{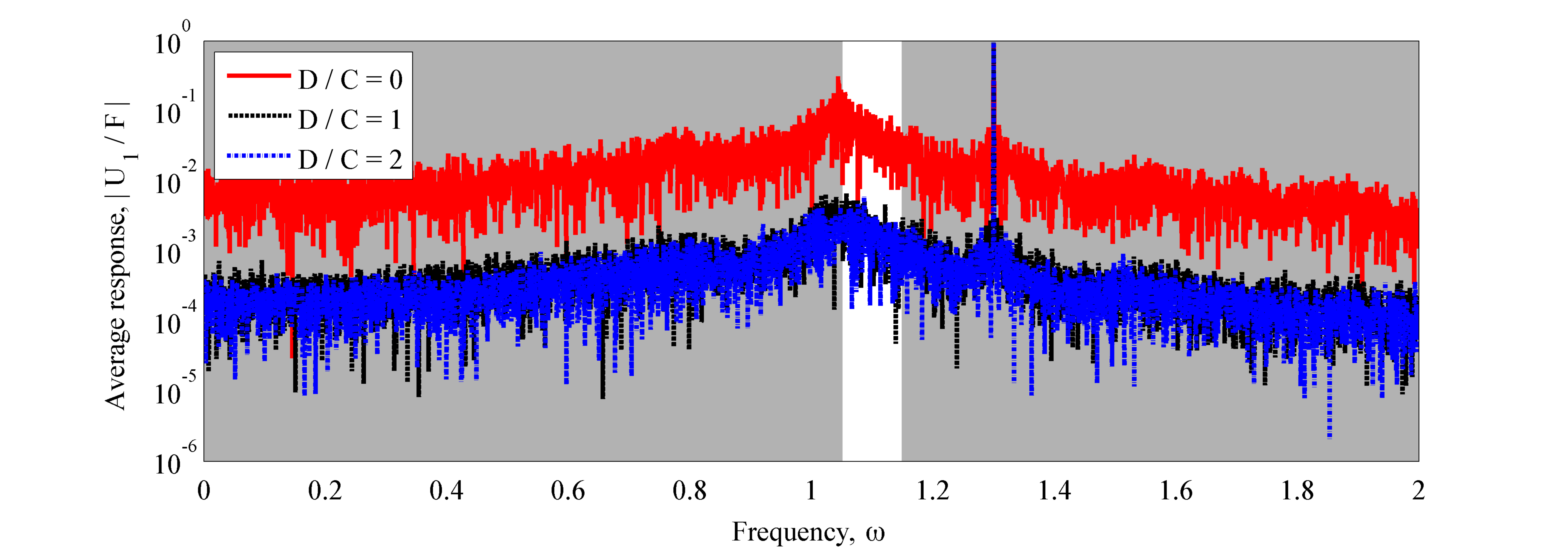}
	\caption{Average frequency spectra of the driven unit for the ordered and disordered structures. The common peak at 1.30 corresponds to the forcing frequency ($\Omega=1.30$). The grey area denotes the linear stop band. 
	%The white background corresponds to the frequency range between the first and last linear natural frequencies of the ordered structure.
	}
	\label{fig:U1spect}
\end{figure}

\begin{figure}[hbt]
\centering
	\includegraphics[width=\linewidth]{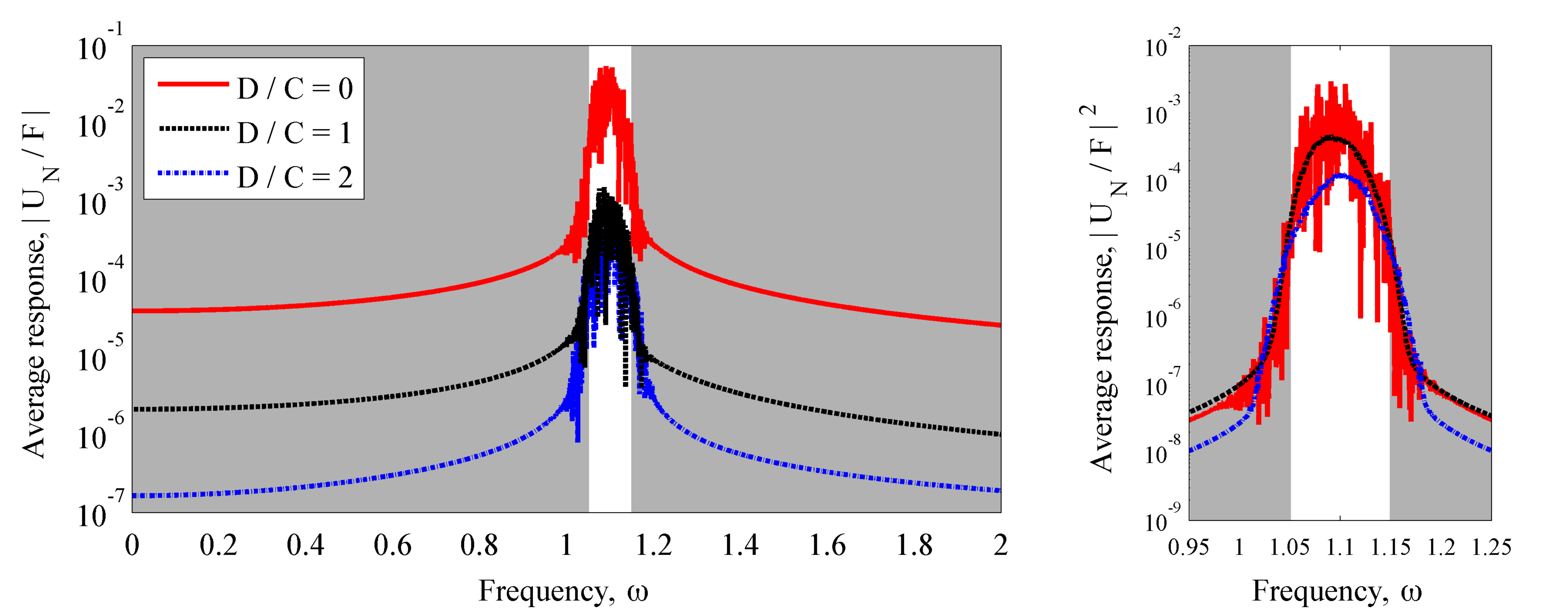}
	\caption{Average transmitted spectra (at $n=N$) for the ordered and disordered structures; (a) average spectra based on complex-valued amplitudes, (b) average squared spectra in the vicinity of linear pass band. The average transmitted spectra lie within the linear pass band. Less energy is transmitted above the supratransmission threshold as we increase the strength of disorder. 
	%The white background corresponds to the frequency range between the first and last linear natural frequencies of the ordered structure.
	The grey area denotes the linear stop band. 
	}
	\label{fig:UNspect}
\end{figure}

%As the periodic solutions lose their stability through the saddle-node bifurcation, the response moves to an anharmonic branch. 
For the system parameters used in this work, we have found that the post-threshold branch within the stop band is chaotic, with a frequency spectrum similar to Figure \ref{fig:supra0spec}. 
To understand the average influence of disorder on the transmitted waves above the threshold, we compare average frequency spectra of the first and last units in the structure. These results are obtained for the same ensembles as in Section \ref{sec:Eprofiles}. For disordered systems, the frequency spectrum of each individual realization is obtained as explained in Section \ref{sec:supra}. Before averaging the frequency spectra within each ensemble, the individual complex-valued spectra for each realization is normalized with its forcing amplitude. This is because the threshold force is different for each individual realization (recall Figure \ref{fig:examples}). 
%We have normalized each of these complex-valued spectra with their corresponding forcing amplitude and then taken their arithmetic mean to obtain the average frequency spectrum for an ensemble.  

Figure \ref{fig:U1spect} shows the average frequency spectra of the driven unit ($n=1$) for the ordered and disordered structures. 
The three spectra have a common pronounced peak at 1.30, which corresponds to the forcing frequency ($\Omega = 1.30$). This is the only dominant frequency component for the two disordered structures. At other frequencies, it is not easy to distinguish between the two disordered spectra, though they both have much lower amplitudes than those of the ordered structure. 
Based on the dominant peak in the average spectra of the disordered structures, it is tempting to infer that the response of the driven unit is harmonic on average. This is not correct, however; the frequency component for a given realization is similar to that of the ordered structure in terms of overall magnitude. When averaging over an entire ensemble, the phases are incoherent for any given frequency component (other than 1.30). As a result, the average spectrum of an ensemble will have a much lower amplitude than its individual realizations.

Figure \ref{fig:UNspect}(a) shows the average frequency spectra at the end of the structure ($n=N$) for the ordered and disordered structures. 
The most notable feature of the transmitted spectra is that frequency components within the linear stop band (the area with grey background) have significantly decreased compared to $n=1$; cf. Figure \ref{fig:U1spect}. The three spectra are therefore similar in the sense that they contain frequencies predominantly within the linear pass band. 
As the strength of disorder is increased, the amplitudes at different frequency components decrease, most significantly within the stop band. Overall, there is less energy transmitted to the end of the structure as disorder strength increases. This is consistent with the decrease in $E_N$ from Figure \ref{fig:Eprofiles}. 

Once we realize that frequency components of the transmitted waves lie within and near the pass band, we can explain, albeit \emph{qualitatively}, certain aspects of disorder effects from a linear perspective as well. 
Firstly, we expect from linear theory that within the stop band disorder localizes energy near the source of excitation and that less energy is transmitted through the structure as the strength of disorder increases. This is consistent with the average energy profiles in Figure \ref{fig:Eprofiles}(a). It also explains why within the stop band disorder has a more significant influence on the response of the structure above the transmission threshold than below it. 
Below the threshold, the structure behaves linearly and is therefore barely influenced by disorder. Above the threshold, on the other hand, the transmitted waves lie within the pass band, where disorder affects the results most significantly. 

Secondly, we showed for a linear system that the transmitted energy covers a relatively wider frequency range as disorder increases (see Figure~\ref{fig:omegan}). The data in Figure \ref{fig:UNspect}(a) is not conclusive in this regard though. To investigate this point further, we have computed the average squared spectra for each ensemble:
{for each realization within an ensemble, the absolute value of the frequency spectrum is squared, then the arithmetic mean over all realizations is used as the average squared spectrum of the ensemble} -- notice that energy is related to squared amplitudes. 
%Notice that although squaring the spectra removes phase information, they are not required for energy calculations. 
%Notice that although phase information is removed by squaring the spectra, it is not required for energy calculations. 
We have shown the squared spectra in Figure \ref{fig:UNspect}(b) for frequency components close to the pass band. 
We see that as disorder increases, there is less energy in the transmitted waves. Also, energy transmission occurs over a slightly wider frequency range for stronger disorder strengths. This widening of the transmission band is similar to linear systems, but less pronounced. Notice, however, that this widening occurs at very low amplitudes and can therefore pose challenges to experiments. 
%(relative to amplitudes at other frequencies) that it may not be measurable in practice for either linear or nonlinear systems. 

%One could also make predictions of the fate of $E_N$ using Figures \ref{fig:UNspect}(a) and \ref{fig:UNspect}(b), and compare with the results in Figure \ref{fig:Eprofiles}(a). It turns out that predictions base on  Figure \ref{fig:UNspect}(b) are more accurate. We note that the real-valued squared spectra used in Figure \ref{fig:UNspect}(b) are energy-based and do not include phase information. It is therefore not surprising that Figure \ref{fig:UNspect}(b) can explain the trend in $E_N$ more accurately. 

%%%%%%%%%%%%%%%%%%%%%%%%%%%%%%%%%%%%%%%%%%%%%%%%%%%%%%%%%%%%%%%%%%%%%%%%%%%%%%%%%%%%
%%%%%%%%%%%%%%%%%%%%%%%%%%%%%%%%%%%%%%%%%%%%%%%%%%%%%%%%%%%%%%%%%%%%%%%%%%%%%%%%%%%%
\subsection{Prediction of Average Transmitted Energies Based on Linear Theory}
\label{sec:Elinear}

\begin{figure}[hbt]
\centering
	\includegraphics[width=\linewidth]{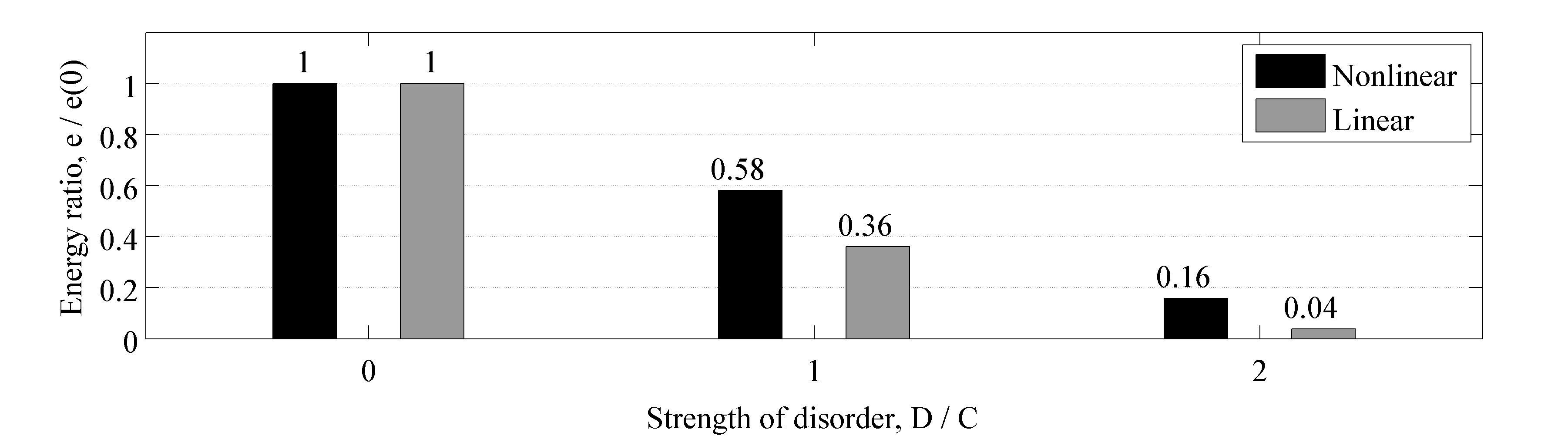}%
	\caption{Normalized transmitted energy ratios, $e/e(0)$, for different values of disorder. The numerical value of each bar is shown above it. These results indicate that linear theory cannot be used for making quantitative predictions of disorder effects above the threshold.  
	}
	\label{fig:ENE1}
\end{figure}

As we discussed in Sections \ref{sec:Eprofiles} and \ref{sec:spectra}, the average behavior of the transmitted waves above the threshold is reminiscent of the average linear behavior of disordered structures. In this light, we ask whether the average transmitted energies above the threshold can be predicted based on linear theory. We introduce the transmitted energy ratio as 
\begin{equation}
	\label{ENE1}
	e = e \left(D/C\right) \equiv \frac{<E_N\left(D/C\right)>}{<E_1\left(D/C\right)>} 
\end{equation}
At a given strength of disorder, $e$ describes the ratio of average transmitted energy to energy in the driven unit. 
We show the normalized transmitted energy ratios $e / e(0)$ for both the linear and nonlinear structures in Figure~\ref{fig:ENE1}. For each system, the transmitted energy ratio is normalized to its value for an ordered structure, $e(0)$. For the linear system, we considered the frequency range between 0 and 2 for energy calculation; widening this frequency range did not change the results. 
Although linear theory can be used to make qualitative predictions of the average behavior of the response above the supratransmission threshold, the results in Figure \ref{fig:ENE1} suggest that they are not appropriate for making quantitative predictions. 

%%%%%%%%%%%%%%%%%%%%%%%%%%%%%%%%%%%%%%%%%%%%%%%%%%%%%%%%%%%%%%%%%%%%%%%%%%%%%%%%%%%%
%%%%%%%%%%%%%%%%%%%%%%%%%%%%%%%%%%%%%%%%%%%%%%%%%%%%%%%%%%%%%%%%%%%%%%%%%%%%%%%%%%%%
%%%%%%%%%%%%%%%%%%%%%%%%%%%%%%%%%%%%%%%%%%%%%%%%%%%%%%%%%%%%%%%%%%%%%%%%%%%%%%%%%%%%
%%%%%%%%%%%%%%%%%%%%%%%%%%%%%%%%%%%%%%%%%%%%%%%%%%%%%%%%%%%%%%%%%%%%%%%%%%%%%%%%%%%%
%%%%%%%%%%%%%%%%%%%%%%%%%%%%%%%%%%%%%%%%%%%%%%%%%%%%%%%%%%%%%%%%%%%%%%%%%%%%%%%%%%%%
%%%%%%%%%%%%%%%%%%%%%%%%%%%%%%%%%%%%%%%%%%%%%%%%%%%%%%%%%%%%%%%%%%%%%%%%%%%%%%%%%%%%
%%%%%%%%%%%%%%%%%%%%%%%%%%%%%%%%%%%%%%%%%%%%%%%%%%%%%%%%%%%%%%%%%%%%%%%%%%%%%%%%%%%%
\section{Prediction of the Onset of Transmission}
\label{sec:prediction}

\subsection{Analytical Estimate of the Transmission Threshold}
\label{sec:formula}

We make two assumptions to develop a theoretical framework for predicting the onset of transmission. 
Results from direct numerical simulations suggest that, to some extent, the periodic structure (i) behaves linearly below the threshold; i.e. for $F<F_{th}$, (ii) behaves nonlinearly predominantly at the first unit. These effects are observed because of the weak coupling between units.  
Based on these simplifications, we linearize the system for $n \ge 2$; i.e. treat the nonlinearity locally. Moreover, we only keep the cubic term of $F_M$ at $n=1$; this is done to keep the analysis tractable by pencil and paper. We call this model the \emph{semi-linear} system. 
A similar approach, in terms of treating the nonlinearity locally, is used in \cite{semilinLangley} in the time domain for modeling the response of drill strings. The analysis presented here is performed in the frequency domain and consequently applies only to the steady-state response of the system. 

In the linear part of the semi-linear model ($2 \le n \le N$), we use a transfer-matrix formulation to find the response of the system. We then use a harmonic approximation of the nonlinear response and find an expression for the onset of instability.  
We present the results from this analysis here. Refer to \ref{sec:tm} and \ref{sec:onset} for derivations.

\begin{figure}[htb]
	\centering
	\includegraphics[width=0.55\linewidth]{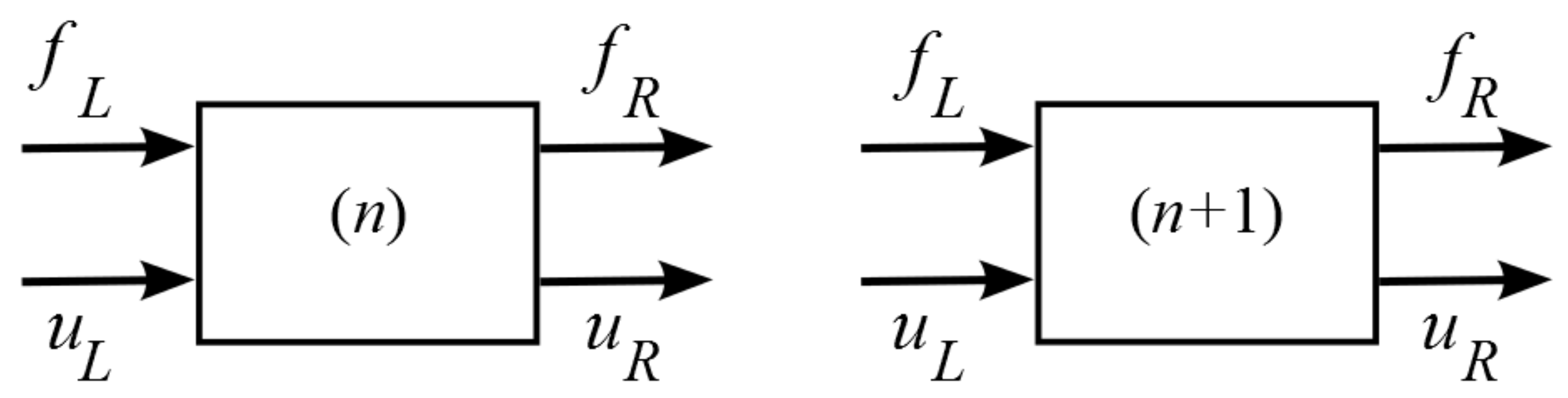}
	\caption{The schematics of two adjacent units in a mono-coupled system.}% The parameters denote complex amplitudes of displacement and forces; i.e. a time dependence of $\exp(\imath \Omega t)$ is assumed. }
	\label{drawing}
\end{figure}

Figure \ref{drawing} shows a schematic representation of two adjacent units in the assembled periodic structure. On either side of each unit (left and right), there is a displacement and force. 
%\footnote{
%The coupling between units occurs through one coordinate of the system -- see the $\Delta^2$ term in \eqref{EOMchain}. The resulting periodic system is accordingly called \emph{mono-coupled}. 
%}. 
We want to relate the force and displacement on the left side of the second unit ($n=2$) to the force and displacement on the right side of the last unit ($n=N$). 
Between adjacent units, we can write
\begin{equation}
	\label{TMunit}
	\vddtwo{u_R}{f_R}^{(n+1)} = [\, T^{(n+1)} \,] \vddtwo{u_R}{f_R}^{(n)}
\end{equation}
where $[\, T^{(n+1)} \,]$ depends on the properties of the $(n+1)$-th unit. The (complex-valued) amplitudes of the displacement and force on the left side of each unit are denoted by $u_L$ and $f_L$, where a time dependence of $\exp(\imath \Omega t)$ is assumed. For each linear unit, we have
\begin{equation}
	\label{TM2by2}
	[T^{(n)}\,] = 
	\begin{bmatrix}
		1+\sigma_n/\Kappa & 1/\Kappa \\
		\sigma_n    & 1
	\end{bmatrix}
\end{equation}
where
\begin{equation}
	\label{sigma}
	\sigma_n \equiv \omega_n^2 + 2 \imath \zeta \Omega - \Omega^2
\end{equation}
Moving along the finite chain from the last unit all the way back to the first unit we can write
\begin{equation}
	\label{TMtotal}
		\vddtwo{u_R}{f_R}^{(N)} \! \! \! 
		= \mathbf{T}_\text{total} \vddtwo{u_R}{f_R}^{(1)} \! \! \! = % [T \,]^{N-1} \vddtwo{u_R}{f_R}^{(1)} 
		\begin{bmatrix}
		T_{11} & T_{12} \\
		T_{21} & T_{22}
		\end{bmatrix}
		\! \vddtwo{u_R}{f_R}^{(1)}  
\end{equation}
where $\mathbf{T}_\text{total} = [T \,]^{N-1}$ for the ordered system and $\mathbf{T}_\text{total} =  [\, T^{(N)} \,] \times ... \times [\, T^{(2)} \,]$ in the disordered case. 
Moving from the right to left side of the nonlinear unit, as explained in  \ref{sec:onset}, the  forcing amplitude at the onset of transmission, $f_{th}$, can be written as
\begin{equation}
	\label{fth}
	F_{th}^2 = 18 \alpha^2 \left( p \pm \sqrt{q^3} \right) 
\end{equation}
where $p$ and $q$ {depend on the components of $\mathbf{T}_\text{total}$ and $[T^{(1)}\,]$.} % -- see \eqref{eqpq} for details 
Supratransmission occurs when the right-hand side of \eqref{fth} is real and positive. In particular, we need $q>0$; this gives the critical frequency at which the enhanced nonlinear transmission starts. 
Threshold curves predicted by \eqref{fth} are \emph{exact} for a periodic structure (ordered or disordered) that has a cubic nonlinearity at $n=1$ and is otherwise linear. We have verified this by comparing analytical predictions to exact numerical computation of the threshold curves. The two results match very well. {We have not included this comparison for brevity.}

Although our derivation was based on a periodic structure with on-site nonlinearity, we expect it to be valid for structures with inter-site nonlinearity as well, provided that `strain variables' (relative displacements of adjacent units) are used. 
A very similar analysis is performed in \cite{JSV1} based on semi-linear models with one and two degrees of freedom. Those results can be obtained by setting $N=1$ and $N=2$ in the formulation presented above.

% COMMENT ON QUINTIC

%%%%%%%%%%%%%%%%%%%%%%%%%%%%%%%%%%%%%%%%%%%%%%%%%%%%%%%%%%%%%%%%%%%%%%%%%%%%%%%%%%%%
%%%%%%%%%%%%%%%%%%%%%%%%%%%%%%%%%%%%%%%%%%%%%%%%%%%%%%%%%%%%%%%%%%%%%%%%%%%%%%%%%%%%

\subsection{Dependence of Threshold Curves on Nonlinearity: Evaluation of Analytical Estimates}
\label{sec:evaluation}

The main limitations of the current analysis are in (i) {confining the nonlinearity to the first (driven) unit} and (ii) ignoring the higher-order nonlinear terms in $F_M$. The assumption of local nonlinearity is expected to hold for weak coupling (i.e. small $C$) and away from the linear pass band, in particular. Keeping the cubic term is expected to work for weak nonlinearity (i.e. small $|u_L^{(1)}|^2$). 
In this section, we keep the strength of coupling unchanged ($C=0.05$) and explore how the two aspects of nonlinearity mentioned above change the threshold curves in an ordered structure. The influence of coupling strength will be explored in Section \ref{sec:strongCoupling}. 
The same conclusions that we draw for an ordered structure apply to individual realizations of disordered structures as well. In an ensemble-average sense, we showed in Section \ref{sec:averageFW} that transmission thresholds do not change away from the pass band. Thus, we expect the results in this section to carry over to average properties of disordered systems as well. 

We compare the threshold curves in the following four nonlinear systems:
\begin{enumerate}[(A)]
	\item Full nonlinearity, global: the system defined in \eqref{EOMchain}.
	\item Cubic nonlinearity, global: the system defined in \eqref{EOMchain}, $F_{M,n}$ truncated at cubic term.
	\item Full nonlinearity, local: the system defined in \eqref{EOMchain}, nonlinear terms of $F_{M,n}$ ignored for $2 \le n \le N$.
	\item Cubic nonlinearity, local: the system defined in \eqref{EOMchain}, $F_{M,n}$ truncated at cubic term, nonlinear terms of $F_{M,n}$ ignored for $2 \le n \le N$. 
\end{enumerate}
We consider both softening and hardening systems -- we note that supratransmission occurs below the pass band in a softening system and above it in a hardening system \cite{JSV1}.
We choose system parameters such that the nonlinear terms in the softening and hardening systems have the same magnitude but opposite signs. The results presented in this section are obtained numerically. 

\begin{figure}[htb]
	\centering
	\includegraphics[width=\linewidth]{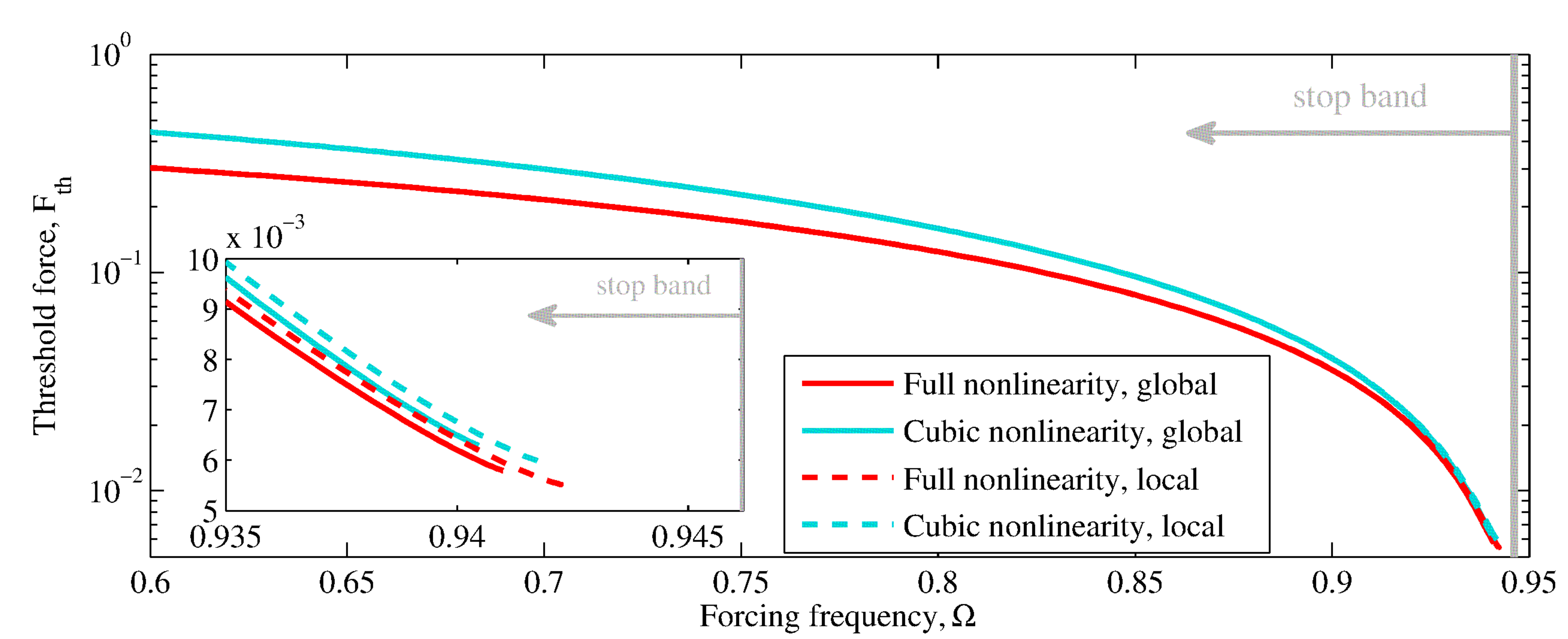}
	\caption{
	Threshold curves for different softening systems with $k_3=-0.2$. 
	Solid curves correspond to systems in which all units are nonlinear (global nonlinearity), dashed curves to systems in which only the first unit is nonlinear (local nonlinearity). Red curves are used when the nonlinear term $F_M$ is used as defined in \eqref{FMn}. Cyan curves are used when $F_M$ is truncated at its cubic term. 
	The edge of the pass band is denoted by the horizontal arrows. The inset shows frequencies close to the pass band edge. 
	}
	\label{fig:FW_soft}
\end{figure}

Figure \ref{fig:FW_soft} shows the threshold curves for the four nonlinear systems for a softening system with cubic coefficient $k_3 = -0.2$. 
Comparing systems with global nonlinearity to those with local nonlinearity, we see that treating the nonlinearity locally results in \revtextt{a slight} overestimation of the threshold force. Nevertheless, this difference is only noticeable at frequencies where the threshold curve starts, which is very close to the pass band (this point is a cusp in the $F-\Omega$ plane -- see Figure~\ref{fig:supra0FWcurve}). The reason for the good agreement between locally- and globally-nonlinear systems is that the coupling between units is very weak ($C \ll 1$). 
As $\Omega$ moves farther into the stop band, we see that a larger force $F_{th}$ is required to trigger instability. As a result of this, the amplitude of vibrations in the driven unit (i.e. $|u_L^{(1)}|$) becomes larger at the onset of instability. This makes the contribution from higher-order nonlinear terms more significant; therefore, the approximation in truncating $F_{M,n}$ at its cubic term becomes less accurate. This is why the difference between threshold curves with full nonlinearity and cubic nonlinearity increases as we move away from the pass band. The same reasoning explains why the system with cubic nonlinearity overestimates the threshold force.

\begin{figure}[htb]
	\centering
	\includegraphics[width=\linewidth]{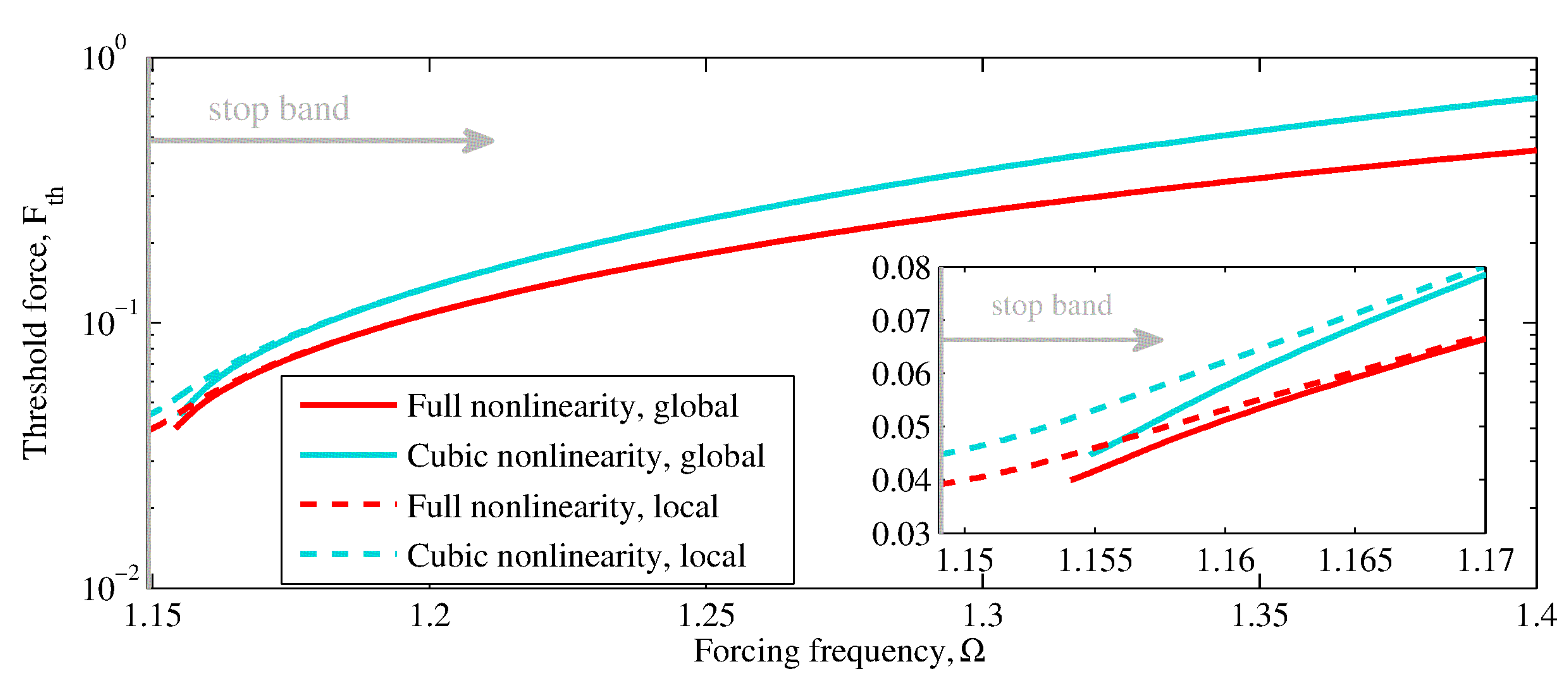}
	\caption{
	Threshold curves for different hardening systems with $k_3=+0.2$. 
	Solid curves correspond to systems in which all units are nonlinear (global nonlinearity), dashed curves to systems in which only the first unit is nonlinear (local nonlinearity). Red curves are used when the nonlinear term $F_M$ is used as defined in \eqref{FMn}. Cyan curves are used when $F_M$ is truncated at its cubic term. 
	The edge of the pass band is denoted by the horizontal arrows. The inset shows frequencies close to the pass band edge. 
	}
	\label{fig:FW_hard}
\end{figure}

Truncating the nonlinearity at its cubic term or restricting it to $n=1$ has the same qualitative effect on threshold curves in hardening systems as it does in softening systems. 
We can see this in Figure \ref{fig:FW_hard}, which shows the four threshold curves for a hardening system with cubic coefficient $k_3=0.2$. 
%The same discussion is valid in comparing the four threshold curves for the hardening system. We have shown these curves in Figure \ref{fig:FW_hard} for a system with cubic coefficient $k_3=0.2$. Comparing the threshold curves in Figures \ref{fig:FW_soft} and \ref{fig:FW_hard}, we notice that truncating the nonlinear term at its cubic term or restricting it to $n=1$ has the  
Apart from these similarities, we see two main differences between threshold curves in hardening and softening systems. Firstly, restricting nonlinearity to the first unit extends the threshold curve to the pass band for a hardening system (compare solid curves to dashed curves in Figure \ref{fig:FW_hard}). To explain this, we recall from Section \ref{sec:examples} that threshold curves trace the locus of saddle-node bifurcations, and from Section \ref{sec:supraPassband} that the solutions within the pass band usually lose stability through a Neimark-Sacker bifurcation. % as the first turning points of NLRMs. 
When the threshold curve of a locally-nonlinear system continues inside the pass band, it means that the corresponding linear solutions lose their stability via saddle-node bifurcation. Thus, treating the nonlinearity locally can predict an incorrect instability mechanism for a globally-nonlinear system close to (and within) the pass band. Further discussion of the exact bifurcation structure in the vicinity of the pass band, and its dependence on damping, is beyond the scope of this work. This is also the frequency range where supratransmission is of less interest, as discussed in Section~\ref{sec:supraDamping}.
%We note, however, that $F_{th}$ does not go to zero at the edge of the pass band for either locally- or globally-nonlinear systems; this occurs because of damping effects~\cite{JSV1}. 

Secondly, approximating the nonlinear term as cubic gives a more accurate estimation of the threshold curves in the softening system than the hardening system.
We can see this by comparing threshold curves in Figures~\ref{fig:FW_soft} and \ref{fig:FW_hard}. For a given distance from pass band, notice that the values of threshold force $F_{th}$ are higher in the hardening system than the corresponding ones in the softening system. 
Thus, just before the onset of instability, the amplitudes of motion are higher in the hardening system. As a result, the higher-order nonlinear terms are more important in determining the onset of transmission in the hardening structure than in the softening one. 
We have confirmed this by computing the threshold curves for systems in which $F_{M,n}$ is truncated at its quintic term (not shown). The threshold curves for the systems with quintic nonlinearity were much closer to the threshold curves of the fully nonlinear system.

%%%%%%%%%%%%%%%%%%%%%%%%%%%%%%%%%%%%%%%%%%%%%%%%%%%%%%%%%%%%%%%%%%%%%%%%%%%%%%%%%%%%
%%%%%%%%%%%%%%%%%%%%%%%%%%%%%%%%%%%%%%%%%%%%%%%%%%%%%%%%%%%%%%%%%%%%%%%%%%%%%%%%%%%%

\subsection{Threshold Curves in Structures with Strong Coupling: Limits of Locally-Nonlinear Behavior}
\label{sec:strongCoupling}

\begin{figure}[htb]
	\centering
	\includegraphics[width=\linewidth]{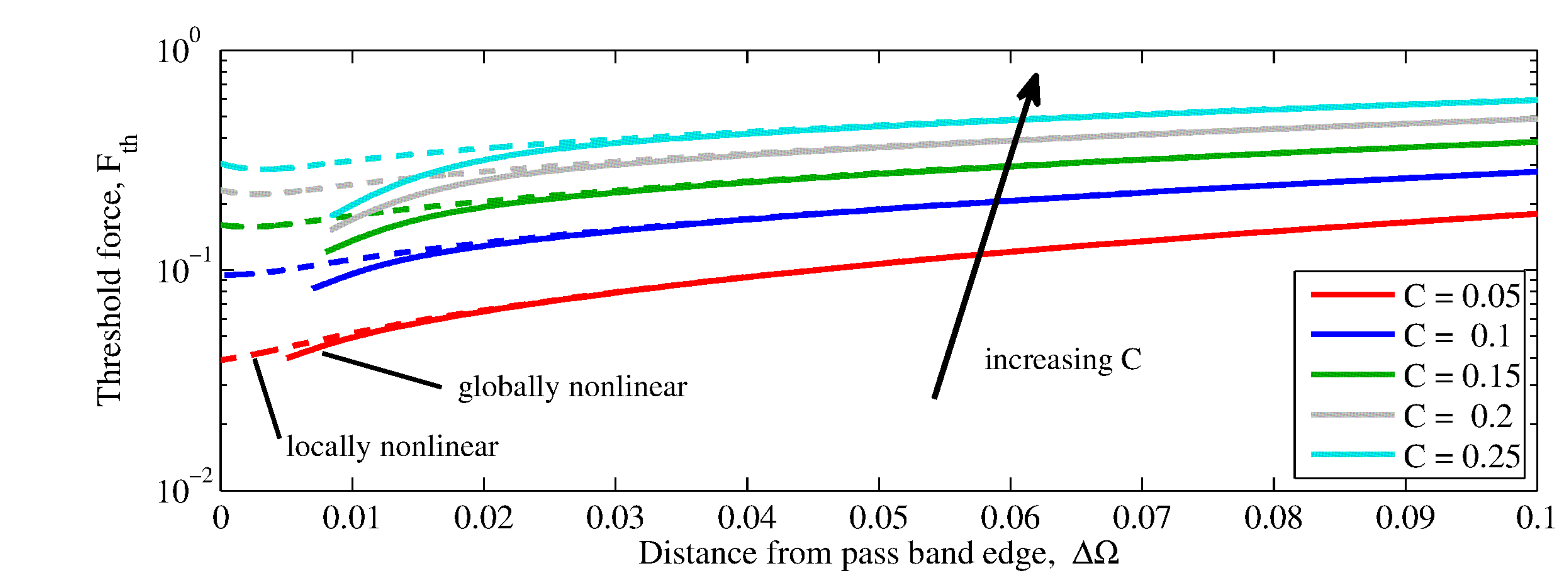}
	\caption{
	Influence of coupling strength on threshold curves for the system with full nonlinearity. Solid curves correspond to the globally nonlinear case and dashed curves to the locally nonlinear case. 
	The horizontal axis shows the distance from pass band edge, where $\Delta \Omega$ is defined in \eqref{deltaOmega}.
	The arrow indicates the direction of increasing coupling strength. As the strength of coupling increases, restricting the nonlinear forces to $n=1$ is inaccurate over a larger frequency range. The red curves ($C=0.05$) are reproduced from Figure \ref{fig:FW_hard}. 
	}
	\label{fig:FWcoupl}
\end{figure}

As already stated, the basis for treating the nonlinearity locally is weak strength of coupling between units. This approximation is expected to lose accuracy as the strength of coupling increases. 
To investigate this, we compute the threshold curves at different values of $C$ for two hardening structures with full nonlinearity; i.e. having the complete nonlinear form of $F_{M,n}$ from~\eqref{FMn}. In one of them nonlinearity is treated locally, while in the other one all units are nonlinear (globally nonlinear). We compare these threshold curves in Figure \ref{fig:FWcoupl} as a function of the distance from pass band, $\Delta \Omega$. This is because increasing $C$ moves the pass band edge to higher frequencies. Thus, we define 
%$\Delta \Omega \equiv \Omega - \omega_{N}$, 
 \begin{equation}
 	\label{deltaOmega}
 	\Delta \Omega \equiv \Omega - \omega_{N}
 \end{equation}
where $\omega_N=\omega_N(C)$ is the largest linear natural frequency of the structure. 

We see in Figure \ref{fig:FWcoupl} that the threshold curves extend to the pass band for the locally nonlinear structures (see the discussion in Section \ref{sec:evaluation}). There is therefore a frequency range in the vicinity of the pass band where using a locally-nonlinear model predicts an incorrect instability mechanism; this frequency range increases with the strength of coupling. 
Furthermore, the assumption of local nonlinearity becomes more inaccurate as $C$ increases, and overestimates the threshold curves over a larger frequency range. 
%Figure \ref{fig:FWcoupl} shows that as $C$ increases, the assumption of local nonlinearity overestimates the threshold curves over a larger frequency range. In particular, treating nonlinearity locally is inaccurate close to the pass band. 
For frequencies far from the pass band, restricting nonlinearity to the driven unit gives an accurate prediction of the onset of supratransmission. The reason is that, in this frequency range, the linear response is highly localized to the driven unit and dispersion (Bragg scattering) dominates over nonlinear forces. As $C$ increases, the assumption of local nonlinearity is accurate over a smaller frequency range. 

%the linear response is highly localized to the driven unit and Bragg scattering dominates over nonlinear forces. In this frequency range, the assumption of local nonlinearity introduces little error in estimating the threshold force. 

%%%%%%%%%%%%%%%%%%%%%%%%%%%%%%%%%%%%%%%%%%%%%%%%%%%%%%%%%%%%%%%%%%%%%%%%%%%%%%%%%%%%
%%%%%%%%%%%%%%%%%%%%%%%%%%%%%%%%%%%%%%%%%%%%%%%%%%%%%%%%%%%%%%%%%%%%%%%%%%%%%%%%%%%%
%%%%%%%%%%%%%%%%%%%%%%%%%%%%%%%%%%%%%%%%%%%%%%%%%%%%%%%%%%%%%%%%%%%%%%%%%%%%%%%%%%%%
%%%%%%%%%%%%%%%%%%%%%%%%%%%%%%%%%%%%%%%%%%%%%%%%%%%%%%%%%%%%%%%%%%%%%%%%%%%%%%%%%%%%
%%%%%%%%%%%%%%%%%%%%%%%%%%%%%%%%%%%%%%%%%%%%%%%%%%%%%%%%%%%%%%%%%%%%%%%%%%%%%%%%%%%%
%%%%%%%%%%%%%%%%%%%%%%%%%%%%%%%%%%%%%%%%%%%%%%%%%%%%%%%%%%%%%%%%%%%%%%%%%%%%%%%%%%%%
%%%%%%%%%%%%%%%%%%%%%%%%%%%%%%%%%%%%%%%%%%%%%%%%%%%%%%%%%%%%%%%%%%%%%%%%%%%%%%%%%%%%
%%%%%%%%%%%%%%%%%%%%%%%%%%%%%%%%%%%%%%%%%%%%%%%%%%%%%%%%%%%%%%%%%%%%%%%%%%%%%%%%%%%%
%%%%%%%%%%%%%%%%%%%%%%%%%%%%%%%%%%%%%%%%%%%%%%%%%%%%%%%%%%%%%%%%%%%%%%%%%%%%%%%%%%%%
%%%%%%%%%%%%%%%%%%%%%%%%%%%%%%%%%%%%%%%%%%%%%%%%%%%%%%%%%%%%%%%%%%%%%%%%%%%%%%%%%%%%
%%%%%%%%%%%%%%%%%%%%%%%%%%%%%%%%%%%%%%%%%%%%%%%%%%%%%%%%%%%%%%%%%%%%%%%%%%%%%%%%%%%%
%%%%%%%%%%%%%%%%%%%%%%%%%%%%%%%%%%%%%%%%%%%%%%%%%%%%%%%%%%%%%%%%%%%%%%%%%%%%%%%%%%%%
\section{Concluding Remarks}
\label{sec:conclude}

%We have presented a theoretical study on the statistical effects of linear uniform disorder on supratransmission. 
We have studied the competition between dispersion effects and nonlinearity in the context of the supratransmission phenomenon in discrete nearly-periodic (disordered) structures. 
We considered a damped nonlinear periodic structure of finite length with weakly-coupled units. 
We showed that although individual realizations of a disordered structure have different onsets of supratransmission, the threshold curve is robust to disorder when averaged over an entire ensemble. For harmonic excitation away from the pass band edge, increasing the strength of disorder has negligible influence on transmitted energies below the onset of supratransmission. In contrast, we found average transmitted energies to decrease with disorder above the transmission threshold. 
This happens because the average frequency spectrum of transmitted waves lies within the linear pass band of the structure, where disorder is known to localize the response to the driven unit (Anderson localization). Overall, the influence of disorder decreases as the forcing frequency moves away from the pass band and supratransmission force threshold increases.

We provided analytical estimates for predicting the onset of transmission for weakly-coupled structures. This formulation is exact for a disordered periodic structure that has cubic nonlinearity in its driven unit and is linear otherwise. We further studied the range of validity of the analysis by studying the dependence of threshold curves on nonlinearity and strength of coupling. 
For forcing frequencies away from the pass band edge, where the linear solution is highly localized, we found that the nonlinear forces are confined to the driven unit. In this frequency range, truncating the nonlinear forces at their cubic terms results in overestimation of the onset of supratransmission; however, using the complete form of the nonlinear forces gives a very accurate prediction of the threshold curves. 
Closer to the pass band edge, the linear response is no longer highly localized and the nonlinearity spreads to other units.  Strong coupling invalidates our analysis, but is of less interest in the context of disorder effects in periodic structures. 
%Truncating the nonlinear forces or treating them locally results in overestimation of the threshold curve. 

Experiments are in progress to validate the findings of this study. \revtext{Throughout this work, we focused on supratransmission in the following sense: loss of stability of periodic solutions at the first turning point of the nonlinear response manifold~\cite{supra_aubry}. 
Nevertheless, we showed that damping may bring about other stable periodic solutions at the first turning point. In the structure studied in this work, supratransmission could still take place at subsequent turning points. Further numerical studies of this feature remains to be done.}
While the present study focused entirely on tonal excitation, supratransmission phenomenon arising from band-limited excitation remains as an open problem. %As well, experiments are in progress to validate the findings of this study. 

%%%%%%%%%%%%%%%%%%%%%%%%%%%%%%%%%%%%%%%%%%%%%%%%%%%%%%%%%%%%%%%%%%%%%%%%%%%%%%%%%%%%
%%%%%%%%%%%%%%%%%%%%%%%%%%%%%%%%%%%%%%%%%%%%%%%%%%%%%%%%%%%%%%%%%%%%%%%%%%%%%%%%%%%%
%%%%%%%%%%%%%%%%%%%%%%%%%%%%%%%%%%%%%%%%%%%%%%%%%%%%%%%%%%%%%%%%%%%%%%%%%%%%%%%%%%%%
%%%%%%%%%%%%%%%%%%%%%%%%%%%%%%%%%%%%%%%%%%%%%%%%%%%%%%%%%%%%%%%%%%%%%%%%%%%%%%%%%%%%
%%%%%%%%%%%%%%%%%%%%%%%%%%%%%%%%%%%%%%%%%%%%%%%%%%%%%%%%%%%%%%%%%%%%%%%%%%%%%%%%%%%%
%%%%%%%%%%%%%%%%%%%%%%%%%%%%%%%%%%%%%%%%%%%%%%%%%%%%%%%%%%%%%%%%%%%%%%%%%%%%%%%%%%%%
%%%%%%%%%%%%%%%%%%%%%%%%%%%%%%%%%%%%%%%%%%%%%%%%%%%%%%%%%%%%%%%%%%%%%%%%%%%%%%%%%%%%
%%%%%%%%%%%%%%%%%%%%%%%%%%%%%%%%%%%%%%%%%%%%%%%%%%%%%%%%%%%%%%%%%%%%%%%%%%%%%%%%%%%%
%%%%%%%%%%%%%%%%%%%%%%%%%%%%%%%%%%%%%%%%%%%%%%%%%%%%%%%%%%%%%%%%%%%%%%%%%%%%%%%%%%%%
%%%%%%%%%%%%%%%%%%%%%%%%%%%%%%%%%%%%%%%%%%%%%%%%%%%%%%%%%%%%%%%%%%%%%%%%%%%%%%%%%%%%
%%%%%%%%%%%%%%%%%%%%%%%%%%%%%%%%%%%%%%%%%%%%%%%%%%%%%%%%%%%%%%%%%%%%%%%%%%%%%%%%%%%%
%%%%%%%%%%%%%%%%%%%%%%%%%%%%%%%%%%%%%%%%%%%%%%%%%%%%%%%%%%%%%%%%%%%%%%%%%%%%%%%%%%%%
\section*{Acknowledgment}

\revtextt{We thank Magnus Johansson and George Kopidakis for helpful discussions.}
We acknowledge financial support from the Canada Research Chairs program; Natural Sciences and Engineering Research Council (NSERC), Canada, through the Discovery Grant program; Canada Foundation for Innovation (CFI). B.Y. acknowledges the support of the University of British Columbia via a 4YF fellowship.



%%%%%%%%%%%%%%%%%%%%%%%%%%%%%%%%%%%%%%%%%%%%%%%%%%%%%%%%%%%%%%%%%%%%%%%%%%%%%%%%%%%%
%%%%%%%%%%%%%%%%%%%%%%%%%%%%%%%%%%%%%%%%%%%%%%%%%%%%%%%%%%%%%%%%%%%%%%%%%%%%%%%%%%%%
%%%%%%%%%%%%%%%%%%%%%%%%%%%%%%%%%%%%%%%%%%%%%%%%%%%%%%%%%%%%%%%%%%%%%%%%%%%%%%%%%%%%
%%%%%%%%%%%%%%%%%%%%%%%%%%%%%%%%%%%%%%%%%%%%%%%%%%%%%%%%%%%%%%%%%%%%%%%%%%%%%%%%%%%%
%%%%%%%%%%%%%%%%%%%%%%%%%%%%%%%%%%%%%%%%%%%%%%%%%%%%%%%%%%%%%%%%%%%%%%%%%%%%%%%%%%%%
%%%%%%%%%%%%%%%%%%%%%%%%%%%%%%%%%%%%%%%%%%%%%%%%%%%%%%%%%%%%%%%%%%%%%%%%%%%%%%%%%%%%
\appendix       %%% starting appendix

%%%%%%%%%%%%%%%%%%%%%%%%%%%%%%%%%%%%%%%%%%%%%%%%%%%%%%%%%%%%%%%%%%%%%%%%%%%%%%%%%%%%
%%%%%%%%%%%%%%%%%%%%%%%%%%%%%%%%%%%%%%%%%%%%%%%%%%%%%%%%%%%%%%%%%%%%%%%%%%%%%%%%%%%%
\section{Comparison of Decay Exponents}%\hspace{15mm}
\label{sec:xicompare}

\begin{figure}[hbt]
\centering
	\includegraphics[height=5.5cm]{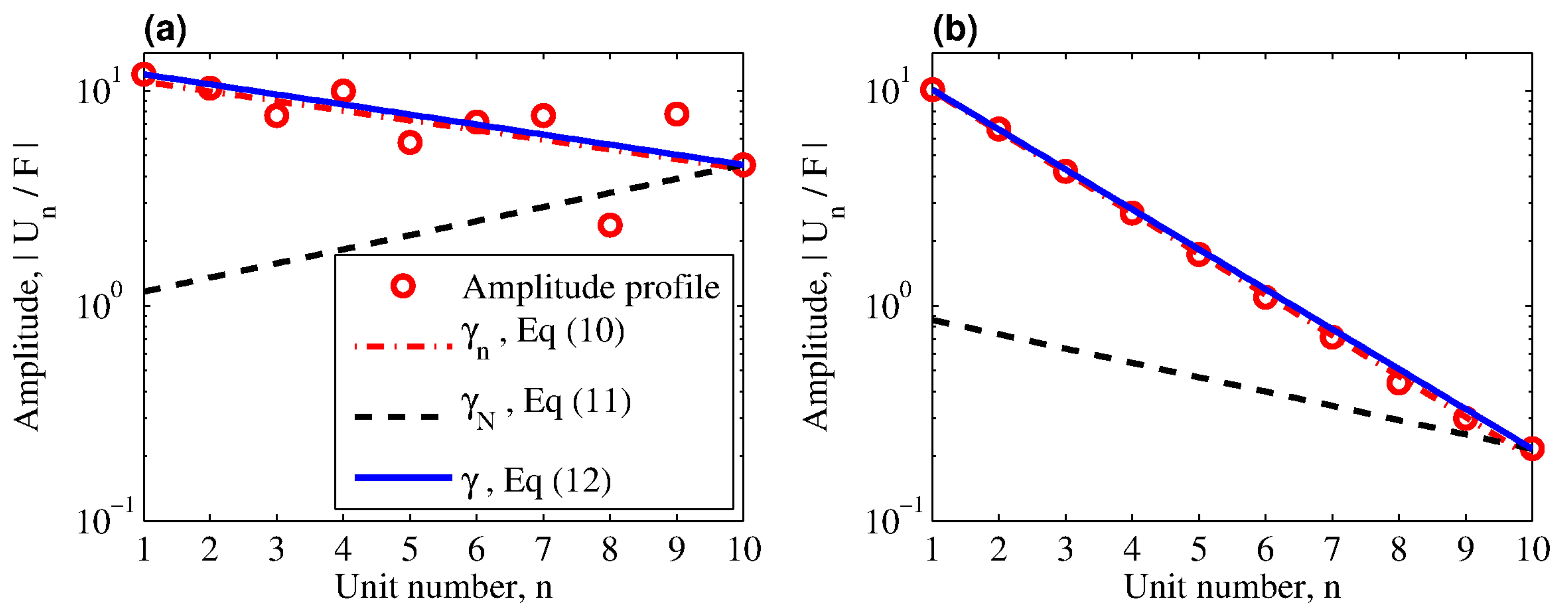}
	\caption{Comparison of the three decay exponents in describing amplitude profiles; (a) an extended response, (b) a localized response. The red circles denote the actual response at each unit, red dash-dotted lines are obtained based on $\gamma_n$ (curve fit), black dashed lines are based on $\gamma_N$ and blue solid lines are based on $\gamma$. }
	\label{fig:xicompare}
\end{figure}

Here, we compare the three decay exponents $\gamma_n$, $\gamma_N$ and $\gamma$ that are defined in \eqref{xiMain}, \eqref{xiHodges} and \eqref{xiUNU1}, respectively. We use two different amplitude profiles for this purpose. These are the average amplitude profiles for the linear system considered in Section \ref{sec:disordered} at $\Omega = 1.12$ for $D/C=0$ (extended response) and $D/C=2$ (localized response), reproduced from Figure \ref{fig:anderson_Un}. We show these amplitude profiles in Figure \ref{fig:xicompare}, along with the estimates obtained based on $\gamma_n$, $\gamma_N$ and $\gamma$. We see that the proposed definition of the decay exponent ($\gamma$) describes the actual amplitude profile better than the classical definition ($\gamma_N$). The advantage of using $\gamma$ over a decay exponent based on curve fitting (i.e. $\gamma_n$) is that the former can be used for obtaining analytical estimates of the response. %Since we do not make use of these analytical expressions in this work, we do not further elaborate on them. 

\begin{figure}[hbt]
\centering
	\includegraphics[height=5.5cm]{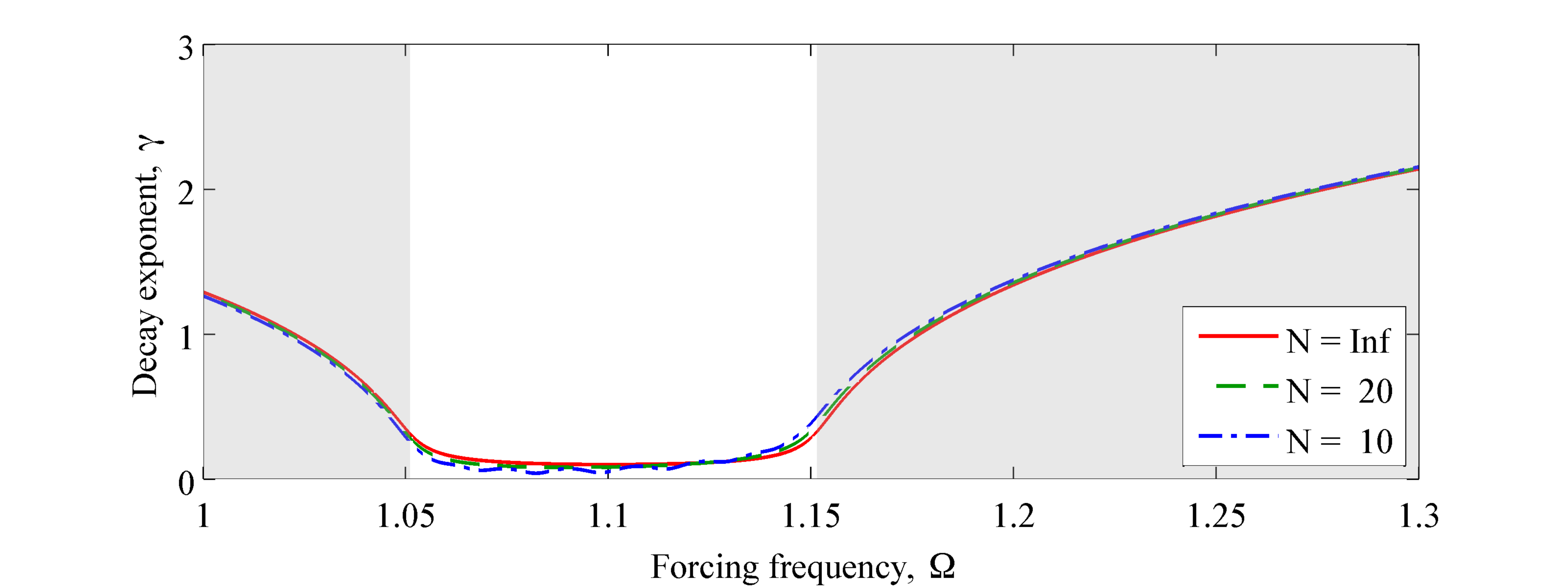}%{old_decayExpo_01_length.png}
	\caption{
	Dependence of the decay exponent on the number of units, $N$, for a linear ordered structure. As the size of the system increases, the decay exponent of the finite system approaches that of the infinite system; i.e. $\gamma \rightarrow \gamma_0$ as $N \rightarrow \infty$.
	The grey area corresponds to the stop band. 
	%The white background corresponds to the frequency range between the first and last linear natural frequencies of the ordered structure.
	}
	\label{fig:xiNinfty}
\end{figure}

Figure \ref{fig:xiNinfty} shows the influence of the number of units on the decay exponent $\gamma$ in an ordered structure; the results for $N=10$ are reproduced from Figure \ref{fig:omegan}(a). We see that as $N$ increases, the results for a finite system approach those for the infinite system based on $\gamma_0$ defined in \eqref{bandLin}. 
It is possible to show analytically for an ordered structure that in fact $\gamma \rightarrow \gamma_0$ as $N \rightarrow \infty$. 

%system('gs -o -q -sDEVICE=png256 -dEPSCrop -r300 -omcmat_anderson_Un.png mcmat_anderson_Un.eps')

%%%%%%%%%%%%%%%%%%%%%%%%%%%%%%%%%%%%%%%%%%%%%%%%%%%%%%%%%%%%%%%%%%%%%%%%%%%%%%%%%%%%
%%%%%%%%%%%%%%%%%%%%%%%%%%%%%%%%%%%%%%%%%%%%%%%%%%%%%%%%%%%%%%%%%%%%%%%%%%%%%%%%%%%%
\section{Derivation of the Transfer Matrix Formulation}%\hspace{15mm}
\label{sec:tm}

% [to be double-checked for consistency of notation] $\alpha \rightarrow k_3$
% \vspace{3mm}

\begin{figure}[bht]
	\centering
	\includegraphics[width=.3\linewidth]{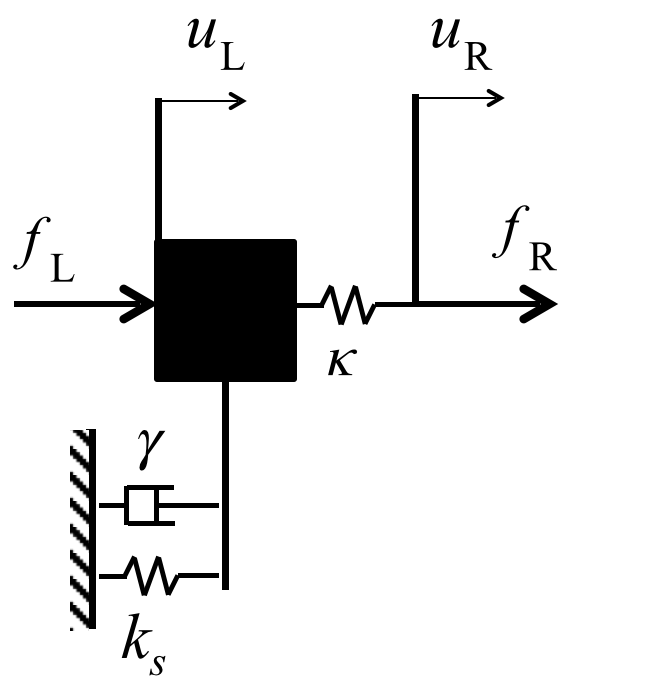}
	\caption{The schematic of a unit cell. We have $k_s=\omega_n^2$ for linear units, while $k_s = \omega_n^2 + 3/4k_3 |u_L^{(n)}|^2$ for nonlinear units.}
	\label{schemaTM}
\end{figure}

Figure \ref{schemaTM} shows a schematic of a unit cell. The unit cell consists of a unit mass, a coupling spring ($\Kappa$), a grounding spring ($k_s$), and a damper  ($2\zeta$). 
Force equilibrium at the left and right ends of the $n$-th unit, respectively, give the following relations
\begin{subequations}
	\label{EOMcell}
	\begin{align}
		\sigma_n u_L^{(n)} + \Kappa (u_L^{(n)} - u_R^{(n)}) &= f_L ^{(n)}\\
		\Kappa (u_R^{(n)} - u_L^{(n)}) &= f_R^{(n)} 
	\end{align}
\end{subequations}
where $\sigma_n$ is defined in \eqref{sigma}. 
From compatibility and force equilibrium between adjacent units we have
\begin{subequations}
	\label{compatibility}
	\begin{align}
		u_L^{(n+1)} &=  u_R^{(n)} \\
		f_L^{(n+1)} &= -f_R^{(n)}
	\end{align}
\end{subequations}
We can rearrange \eqref{EOMcell} to have the `left' variables in terms of the `right' variables. Then, using \eqref{compatibility}, we have
\begin{subequations}
	\begin{align}
		u_R^{(n)} &= (1+\sigma_n/\Kappa) u_R^{(n-1)} + f_R^{(n-1)}/\Kappa \\
		f_R^{(n)} &= \sigma_n u_R^{(n-1)} + f_R^{(n-1)}
	\end{align}
\end{subequations}
which gives the transfer matrix in \eqref{TM2by2}.

% Notice that a free boundary condition is realized by imposing $f_R^{(N)}=0$; see \eqref{EOMcell}.

%%%%%%%%%%%%%%%%%%%%%%%%%%%%%%%%%%%%%%%%%%%%%%%%%%%%%%%%%%%%%%%%%%%%%%%%%%%%%%%%%%%%
%%%%%%%%%%%%%%%%%%%%%%%%%%%%%%%%%%%%%%%%%%%%%%%%%%%%%%%%%%%%%%%%%%%%%%%%%%%%%%%%%%%%
\section{Derivation of the Onset of Transmission}%\hspace{15mm}
\label{sec:onset}

Applying Newton's second law to the right and left nodes of the nonlinear unit at $n=1$, respectively, results in the following equations
\begin{subequations}
	\begin{align}
		f_R^{(1)} &= \Kappa (u_R^{(1)} - u_L^{(1)}) \\
		f_L^{(1)} &= (\sigma_1 + \beta) u_L^{(1)} + \Kappa (u_L^{(1)} - u_R^{(1)})
	\end{align}
\end{subequations}
where we have defined
\begin{equation}
	\label{beta}
	\beta = \beta\left( u_L \right) \equiv 3/4 \, k_3 |u_L|^2
\end{equation}
We rearrange the terms to get 
\begin{subequations}
	\label{EOMcellNL}
	\begin{align}
		u_R^{(1)} &= (1+(\sigma_1+\beta)/\Kappa) u_L^{(1)} - f_L^{(1)}/\Kappa \\
		f_R^{(1)} &= (\sigma_1+\beta) u_L^{(1)} - f_L^{(1)}
	\end{align}
\end{subequations}
using \eqref{EOMcellNL} in combination with \eqref{TMtotal}, we can relate the response on the right of $n=N$ to the response on the left of $n=1$.  The external force applied to the structure corresponds to $f_L^{(1)}=F$ and, without loss of generality, we take $F$ to be real-valued. 
To enforce the free boundary on the right end of the structure we require $f_R^{(N)}=0$. This, after substituting $u_R^{(1)}$ and $f_R^{(1)}$ from \eqref{EOMcellNL} into \eqref{TMtotal}, results in the following 
\begin{equation}
	\left( T_{21} (1+(\sigma+\beta)/\Kappa) + T_{22} (\sigma+\beta) \right) u_L^{(1)} = 
	\left( T_{21}/\Kappa + T_{22} \right) f_L^{(1)}
\end{equation}
which can be used to find the onset of transmission for the semi-linear system. %For this, we would like to find an expression for $|u_L^{(1)}|$ in terms of the system parameters only. 
Setting
\begin{equation}
	\rho \equiv |u_L^{(1)}|^2
\end{equation}
we can obtain the following cubic equation for $\rho$ 
\begin{equation}
	\label{rcubic}
	a_3 \rho^3 + a_2 \rho^2 + a_1 \rho + a_0 = 0
\end{equation}
where
\begin{subequations}
	\label{anCoeff}
	\begin{align}
		a_3 &= (9/16) \, k_3^2 \left| T_{21}/\Kappa + T_{22} \right|^2 > 0 \\
		a_2 &= (3/2) \, k_3 \operatorname{Re}\!\left\{ 
		(T_{21}/\Kappa + T_{22})^\dagger (\sigma(T_{21}/\Kappa + T_{22})+T_{21}) \right\} \\
		%\left( |T_{21}|^2/\Kappa + \operatorname{Re}\{\sigma\} \left| T_{21}/\Kappa + T_{22} \right|^2  + \operatorname{Re}\left\{ T_{21}T^*_{22} \right\} \right) \\ % + (T^*_{21}T_{22}+T_{21}T^*_{22})/2
		a_1 &= |(1+\sigma/\Kappa) T_{21}+ T_{22} |^2 \\ % |\sigma|^2 | T_{21}/\Kappa + T_{22} |^2 + (1+\sigma_r/\Kappa) |T_{21}|^2 + (\sigma T^*_{21}T_{22}+ \sigma^* T_{21}T^*_{22})/2\\
		a_0 &= -F^2 \left| T_{21}/\Kappa + T_{22} \right|^2 < 0
	\end{align}
\end{subequations}
Here, the superscript $\dagger$ denotes complex conjugate. %Also, we have defined $\sigma = \sigma_r + \imath \sigma_i$; see also \eqref{sigma}. 
%Perhaps the expressions for $a_i$ can be written more elegantly in the future.

Equation \eqref{rcubic} is a cubic polynomial with real coefficients and, depending on the relation between its coefficients, may have three real roots or only one. The onset of transmission is where there are three real roots with two of them being equal (i.e. multiple real roots). Such points correspond to saddle-node bifurcations of the Duffing equation \cite{HolmesRand}. We first re-write \eqref{rcubic} as follows
\begin{equation}
	\label{zcubic}
	\rho^3 + c_2 \rho^2 + c_1 \rho + c_0 = 0
\end{equation}
Now we restrict the coefficients of \eqref{zcubic} such that it has at least two equal real roots by setting \cite{handbook}
\begin{equation}
	c_0 = - 2 \left( p \pm \sqrt{q^3} \right)
\end{equation}
where 
\begin{subequations}
	\label{eqpq}
	\begin{align}
		p &= c_2^3/27 - c_1c_2/6  \\
		q &= c_2^2/9 - c_1/3
	\end{align}
\end{subequations}
Using $c_0 = -(4F/3k_3)^2$, the critical forcing amplitude at the onset of threshold, $F_{th}$, can be written as shown in \eqref{fth}.

In the above analysis, we assumed the structure to have free boundaries at both ends. A very similar formulation applies if the boundary on either end of structure is fixed. 
In this case, the formula in \eqref{fth} remains valid but the coefficients $a_n$ in \eqref{anCoeff} need to be updated. 
For a fixed boundary on the driven end of the structure (left side), we have $f_L^{(1)}=F-\Kappa u_L^{(1)}$. To model a fixed boundary on the right end of the structure, the free boundary condition $f_R^{(N)}=0$ should be replaced with $u_R^{(N)}=0$.

% [comment on consistency with \cite{JSV1}]:
% 
% Also, $f_{cr}^2 \propto 1/\alpha$, which demonstrates the dependence of supratransmission on the type of nonlinearity. 

\section*{References}


\bibliographystyle{elsarticle-num} 
\bibliography{refs_j02_rev}

%% \begin{thebibliography}{00}
%% \end{thebibliography}

\end{document}